\documentclass[10pt,journal,final,twocolumn]{IEEEtran}

\usepackage[T1]{fontenc}
\usepackage{graphicx}
\usepackage{amsthm}
\usepackage{amsfonts,amssymb,amsmath}
\usepackage{mathtools}
\usepackage{subcaption}
\usepackage{color}
\usepackage{algorithm, algorithmic, tabularx}
\usepackage{bm}
\usepackage{cite}

\begin{document}
\title{Intelligent Reflecting Surface-Assisted Bistatic Backscatter Networks: Joint Beamforming and Reflection Design}

\author{Xiaolun Jia,~\IEEEmembership{Member,~IEEE,}
		Xiangyun Zhou,~\IEEEmembership{Senior Member,~IEEE,}
		Dusit Niyato,~\IEEEmembership{Fellow,~IEEE,}\newline
        and~Jun Zhao,~\IEEEmembership{Member,~IEEE}

\thanks{This research was undertaken with the assistance of resources from the National Computational Infrastructure (NCI Australia), an NCRIS enabled capability supported by the Australian Government. The work of Jun Zhao was supported in part by the Nanyang Technological University Startup Grant. (\textit{Corresponding author: Xiaolun Jia.})

X. Jia and X. Zhou are with the School of Engineering, The Australian National University, Canberra, Australia (email: \{xiaolun.jia, xiangyun.zhou\}@anu.edu.au). 

D. Niyato and J. Zhao are with the School of Computer Science and Engineering, Nanyang Technological University, Singapore (email: \{dniyato, junzhao\}@ntu.edu.sg).%

Part of this work was presented at IEEE GLOBECOM 2020 \cite{gc}.}}%

\maketitle

\begin{abstract}
Bistatic backscatter communication (BackCom) allows passive tags to transmit over extended ranges, but at the cost of having carrier emitters either transmitting at high powers or being deployed very close to tags. In this paper, we examine how the presence of an intelligent reflecting surface (IRS) could benefit the bistatic BackCom system. We study the transmit power minimization problem at the carrier emitter, where its transmit beamforming vector is jointly optimized with the IRS phase shifts, whilst guaranteeing a required BackCom performance. A unique feature in this system setup is the multiple IRS reflections experienced by signals traveling from the carrier emitter to the reader, which renders the optimization problem highly nonconvex. Therefore, we propose algorithms based on the minorization-maximization and alternating optimization techniques to obtain approximate solutions for the joint design. We also propose low-complexity algorithms based on successive optimization of individual phase shifts. Our results reveal considerable transmit power savings in both single-tag and multi-tag systems, even with moderate IRS sizes, which may be translated to significant range improvements using the original transmit power or a reduction of the reliance of tags on carrier emitters located at close range.
\end{abstract}


\section{Introduction}

\IEEEPARstart{I}{n} recent years, backscatter communication (BackCom) has emerged as a promising solution to improve device lifetime. BackCom devices (or \textit{tags}) communicate by performing passive modulation onto existing radiofrequency (RF) signals, rendering active RF transmission chains unnecessary. The bistatic BackCom architecture, consisting of a carrier emitter (CE) transmitting a sinusoidal signal and a separately located reader, allows extended communication ranges for tags placed close to the CE. Thus, on top of reducing tag power consumption, bistatic BackCom is also ideally suited to applications requiring extended ranges. This aligns with, and is beneficial to aspects of the Internet of Things (IoT) paradigm such as environmental sensing, pervasive monitoring and smart cities, which aim to provide ubiquitous connectivity throughout society.

Various theoretical and experimental studies have been conducted to characterize the achievable range of tags under bistatic BackCom setups. The work in \cite{Kim14} demonstrated extended ranges achievable with a single tag placed close to a CE with only $20$ mW transmit power. The works in \cite{Fas15, Alev17} studied channel coding schemes at the tag and coherent and noncoherent detectors at the reader; while \cite{Shen16} addressed the phase cancellation problem arising from the use of unmodulated carrier signals, resulting in more effective signal reception. However, these works have mostly considered tag-level modifications, and their effects on communication range extension have been mostly incremental, evidenced by the achieved range over time: $130$ m in \cite{Kim14}, and $145$ m and $148$ m in the follow-up works \cite{Fas15, Alev17}, respectively. Moreover, these ranges were achievable only if tags were within $10$ m of the CE, severely limiting the areas over which tags may be placed. Works which consider changes to the system architecture, which could increase both the communication range and CE-tag separation, are currently lacking. To this end, this paper proposes the introduction of an intelligent reflecting surface (IRS) to the bistatic BackCom system for use in the mentioned applications, and examines how such a change on the system architecture level could benefit the range and flexibility of bistatic BackCom.

\subsection{Literature Review}

Recently, IRSs have been highlighted as key enablers for next-generation communication systems, due to their ability to modify the wireless propagation medium to benefit nearby communication links \cite{R2}. An IRS consists of many reflecting elements, each imposing a controllable phase shift on impinging signals. The phase shifts can be jointly optimized to achieve a constructive (or destructive) effect at a receiver \cite{Lia18}. The studies in \cite{Tang19, Ell19} examined the precise path loss scaling in detail, where the signal strength at a receiver was shown to scale proportionally with the IRS surface area \cite{Ozd19}. Thus, with a reasonably-sized IRS, considerable improvement in communication system performance can be realized.

Due to this favorable property, IRSs hold the potential to aid a range of communication systems in both indoor (e.g., buildings and industrial environments) and outdoor (e.g., cellular networks and smart cities) settings \cite{MDRsurvey}. A number of works have studied the optimization of the IRS phase shifts, often jointly with other variables such as transmit beamforming vectors. The work in \cite{Wu19} considered an IRS-aided downlink multiuser network, and proposed semidefinite relaxation (SDR) and alternating optimization (AO) approaches to handle the intractable phase shift optimization problem while jointly performing beamforming optimization. The work in \cite{R3} extended the system model in \cite{Wu19} to an IRS with discrete phase shifts. To reduce the complexity of channel estimation and optimization over many IRS reflectors, a grouping scheme for adjacent IRS elements was proposed in \cite{Yang19}; while \cite{Pan19} extended the joint beamforming and reflection optimization to a multiuser multiple-input multiple output (MIMO) system.

The potential benefits brought by IRSs to conventional communication systems naturally make them suitable candidates to also assist low-power, IoT-type networks. More recently, works on the use of IRSs to facilitate wireless power transfer and cognitive radio have also appeared. The work in \cite{Wu19b} studied the joint active and passive beamforming optimization problem in IRS-assisted simultaneous wireless information and energy transfer (SWIPT) systems; while \cite{Lyu20} considered a wireless-powered communication network (WPCN) where both the IRS and downlink users harvest energy to achieve self-sustainable operation. For improved energy and spectral efficiency, cognitive radio systems assisted by IRSs have also received attention, where the coexistence of primary and secondary systems creates more complex signal paths. The work in \cite{R1} addressed the weighted sum rate maximization problem in an IRS-assisted MIMO cognitive radio system, while \cite{Xu20, Yuan20} considered full-duplex and multi-IRS variants of the cognitive setup, resulting in the IRS needing to balance over the performance of many separate transmissions.

While IRSs are primarily used to assist actively transmitting devices, as seen in the mentioned works, their ubiquitous applicability creates opportunities for them to assist passive transmitters as well. As a technology reliant on external powering signals, BackCom stands to reap significant benefits from potential IRS assistance. Motivated by this, the authors of \cite{Nem20} pioneered the IRS-assisted BackCom system by studying throughput maximization under orthogonal frequency division multiplexing (OFDM) modulated signals. The work in \cite{Zhao20} examined the error performance of an IRS-aided monostatic-like BackCom system without a direct link between tag and reader. The work in \cite{Abe20} proposed channel estimation schemes for a monostatic IRS-assisted BackCom system; while \cite{Far21} experimentally demonstrated the feasibility of an IRS-aided ambient BackCom setup.

\subsection{Motivation and Contributions}

It is noted that most existing works on IRS-aided BackCom systems have focused on the monostatic or ambient architectures, both of which possess short communication ranges. Bistatic BackCom, on the other hand, has a range comparable to some conventional systems. The introduction of an IRS, as a change on the network infrastructure level, would not only realize extended ranges and device lifetimes, but also reduce tags' reliance on CEs in terms of their separation, which translates to more flexible tag deployment and coverage. Given the expected widespread deployment of IRSs on buildings and other structures, and even the appearance of mobile IRSs \cite{Zhang19}, one may reasonably expect the use-cases of IRS to cover the environmental monitoring application (under the scope of smart agriculture \cite{MDRsurvey}) typical of bistatic BackCom systems. The work in \cite{Chen21} explored the IRS phase shift optimization problem under a bistatic-like BackCom setting. However, as we highlight in the sequel, it has not considered the complete signal model, which includes the unique phenomenon of multiple signal reflections at the IRS arising from the reflecting nature of the BackCom device.

In this paper, we present the complete signal model for an IRS-aided bistatic BackCom system for the first time, which accounts for the presence of additional signal paths under line-of-sight scenarios. We present the first results into the extent of CE power consumption reduction through a transmit power minimization problem involving the IRS phase shifts, and quantify the potential backscattering range improvements. The contributions of this paper are as follows:
\begin{itemize}
\item The IRS-aided bistatic BackCom system is introduced, where a separate IRS assists the backscatter communication from the tags to the reader. To our best knowledge, this is the first work to incorporate IRS into bistatic BackCom systems. Different from prior studies on IRS where the signal traveling from the CE to the reader was assumed to undergo only one reflection at the IRS, we highlight a unique feature in that the co-existence of IRS and BackCom tags leads to two significant reflections at the IRS and thus a new signal model.
\item We study the CE transmit power minimization problem for the bistatic BackCom system, subject to the tags' signal-to-noise ratio (SNR) requirements. Due to the nonconvex nature of the problem, we present approximate solutions for the IRS phase shifts in the single-tag case using the minorization-maximization (MM) algorithm. Though suboptimal, we demonstrate considerable transmit power reductions using the MM algorithm, and reveal the IRS phase shift characteristics when balancing between multiple reflections of the same signal.
\item We extend the problem to a multi-tag scenario, and propose an alternating optimization method with a novel successive refinement (SR) approach for individual IRS phase shifts with lower complexity. The scaling behavior of the transmit power is shown to favor scenarios where more tags are present in the system, in terms of effective power per tag.
\end{itemize}

The rest of this paper is organized as follows. Section II introduces the IRS-aided BackCom system and the signal model. Section III presents the general form of the transmit power minimization problem. Section IV proposes solutions to a base case of the problem with one semi-passive tag and a single-antenna reader. Section V presents the algorithms required to solve the general problem involving multiple passive or semi-passive tags. Numerical results are presented in Section VI and Section VII concludes the paper.

\textit{Notations:} $j = \sqrt{-1}$ denotes the complex unit, and $\mathbb{R}$ and $\mathbb{C}$ denote the set of real and complex numbers, respectively. $\left| \cdot \right|$ and $\mathrm{Re}\{\cdot\}$ denote the magnitude and the real part of a complex number, respectively. $\mathcal{CN}(\mu, \sigma^{2})$ represents a complex Gaussian distribution with mean $\mu$ and variance $\sigma^{2}$. Vector and matrix quantities are denoted using lowercase and uppercase boldface letters, respectively. $\mathbf{I}$ denotes the identity matrix of variable size. $\left\lVert \mathbf{a} \right\rVert$ denotes the Euclidean norm of a vector; $\mathrm{tr}(\mathbf{A})$, $\mathbf{A}^{T}$ and $\mathbf{A}^{H}$ denote the trace, transpose and the Hermitian transpose of $\mathbf{A}$, respectively; and $\odot$ denotes the elementwise product of two vectors or matrices.


\section{System Model}

\subsection{System Setup}

\begin{figure}[!t]
\centering
\includegraphics[width=3.5in]{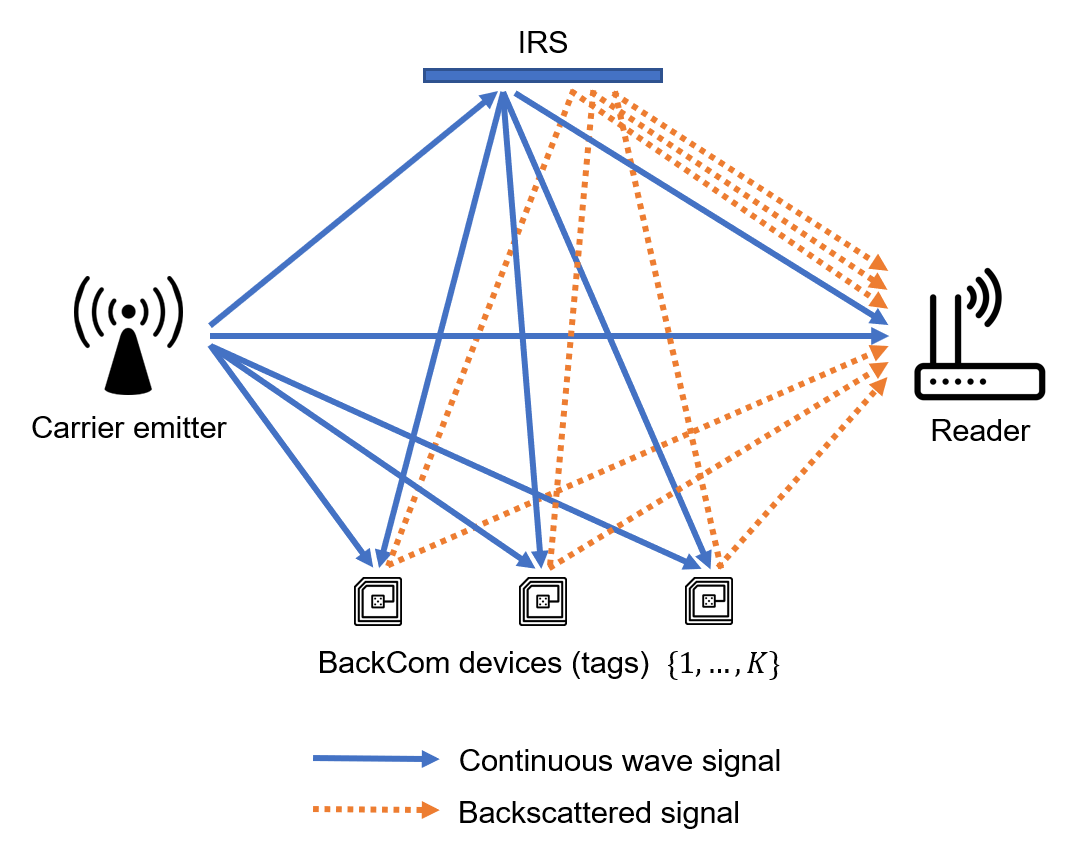}
\caption{The IRS-aided bistatic BackCom system.}
\label{fig_1}
\end{figure}

We consider an IRS-aided bistatic BackCom system in Fig.~\ref{fig_1} with one $L$-antenna CE, $K \geq 1$ single-antenna tags, a reader with $M \geq 1$ antennas, and an IRS with $N$ reflecting elements. Hereafter, the CE, tags, IRS and reader are referred to in subscripts by $C$, $T$, $I$ and $R$, respectively.

We base the system model on a scenario where tags are deployed in the field for an environmental monitoring application. The IRS is either portable or fixed on a building, wall or other structure, and enhances the backscatter communication of the tags. Such a setup can occur in the greenhouse environment considered in \cite{Kam14}, where the IRS may be wall-mounted; or in open environments, where the IRS may be located on the walls of nearby structures such as houses or fences, or even be deployed on-board a hovering UAV to provide localized coverage at a specific time \cite{Zhang19}. The analysis in this paper considers non-negligible channels between all pairs of nodes, in line with experimental works on bistatic BackCom \cite{Kim14, Fas15, Alev17}. For environmental monitoring and related applications, it is typical to install tags at height such that the channels between tags and all other nodes exist. We adopt Rician fading in the channel model, although the choice of fading model does not affect the generality of the analysis in Sections II-V.

The CE transmits a continuous-wave signal with carrier frequency $f_{c}$ to power the tags' communication. Each tag is equipped with two load impedances connected to its antenna, one of which represents an off-state, where the load and antenna impedances are perfectly matched such that the signal is completely absorbed. We consider a generalized tag power supply configuration with a circuit power consumption constraint $\xi$ (in Watts). When $\xi = 0$, the tag is semi-passive, where an on-board battery is the sole energy source for the circuit, and all of the incident signal is used for communication (i.e., backscattered). In the case of $\xi > 0$, a nonzero portion of energy from the incoming signal is used to power the tag.

Tags transmit using generalized binary frequency-shift keying (FSK) modulation, following the assumption in \cite{Kim14, Alev18} for bistatic BackCom systems. Under this scenario, two distinct subcarrier frequencies known by the reader are assigned to tag $k$, and are denoted by $f_{k, 0}$ and $f_{k, 1}$ for bits $0$ and $1$, respectively. That is, the tag switches between its impedances with frequency $f_{k, 0}$ when transmitting bit $0$, and with frequency $f_{k, 1}$ for bit $1$. The frequency-domain representation of the received signal consists of four peaks at $f_{c} \pm f_{k, 0}$ and $f_{c} \pm f_{k, 1}$, which can be resolved using a bank of correlator demodulators at the four frequencies \cite{Kim14}. Due to the small number of tags considered in our setup, we assume that the tags' subcarrier frequencies are distinct enough such that interference between tags' transmissions is negligible. We note that this study aims to conduct initial exploration into the IRS phase~shift design to handle multiple reflections of the same signal, rather than to examine the mitigation of inter-user interference.

Each tag is modeled as a diffuse reflector, which incurs significant signal strength losses. Thus, we ignore signal paths which undergo two or more reflections at tags.\footnote{A reflected signal from an RFID-type tag incurs a near-field path loss of approximately $30$ dB \cite{Dob08}; hence, a signal having undergone an additional reflection is generally weaker than the directly received signal by several orders of magnitude.} However, this assumption does not apply to the reflections at the IRS, particularly for the $C$-$I$-$T$-$I$-$R$ link, with twofold reasoning. First, the strength of the information-bearing signal depends on the continuous-wave signal, both of which are reflected by the IRS. Second, 
the IRS is able to tune its individual elements to balance the signal strengths between the combined CE-tag and tag-reader links\footnote{That is, $C$-$T$ plus $C$-$I$-$T$ links, and $T$-$R$ plus $T$-$I$-$R$ links, respectively, here and elsewhere in the paper.} to maximize the end-to-end signal strength. Therefore, the twice-reflected signals are a unique feature that must be included in the considered system model. 

The IRS is assumed to have its own power supply in order to configure its phase shifts. The reader utilizes linear combining vectors, denoted by $\mathbf{g}_{k} \in \mathbb{C}^{M \times 1}, \ k \in \{1, \ldots, K\}$, to separately decode the information signals from each of the $K$ tags. The set of all combining vectors is denoted by $\mathbf{G} = [\mathbf{g}_{1}, \ldots, \mathbf{g}_{K}] \in \mathbb{C}^{M \times K}$. We extend the reader receiver architecture in \cite{Kim14} to perform combining after demodulation at each antenna, where the aggregate received signal is split into separate components for each tag, such that each tag is assigned its own combiner. 

We assume that the reader has perfect knowledge of all channels in the system. For this exploratory study, using this assumption allows us to characterize the upper bound to system performance. The mathematical treatment and evaluation of channel estimation methods for an IRS-aided bistatic BackCom system is outside the scope of this work. However, any channel estimation method may be intuitively split into phases. In the first phase, the direct $C$-$R$ link can be determined using methods available in the bistatic BackCom literature (e.g., \cite{Fas15}) with the IRS set to the non-reflecting state. Then, the tag is switched on to observe both the $C$-$T$ and $C$-$T$-$R$ channels and to infer the $T$-$R$ channel. In the second phase, the cascaded $C$-$I$-$R$ channel may be resolved into its $C$-$I$ and $I$-$R$ components on the basis of methods from the recent IRS literature \cite{Wei21}. In the third and final phase, the effects of the remaining $I$-$T$ channels can be observed using the information obtained previously on all other signal paths in the system.

\subsection{Signal Model}

Tag $k$ modulates its data symbols onto an incident signal by switching between its impedances to change the power of the reflected signal. The tag's baseband signal is given by $b_{k}(t) = A_{k} - \Gamma_{k}(t)$, where $A_{k} \in \mathbb{C}$ is the antenna structural mode and determines the default amount of signal reflection in the non-reflecting state, and $\Gamma_{k}(t)$ is the reflection coefficient function over time, taking on two possible values $\Gamma_{k, 0}$ and $\Gamma_{k, 1}$, both of unit magnitude or less. The term $A_{k}$ is a constant and can be subtracted from the overall received signal in post-processing. Therefore, as in \cite{Yang18}, we do not take into account the effect of $A_{k}$, and consider the $\Gamma_{k}(t)$ term only.

The energy $\xi_{k}$ required to operate the circuit of tag $k$ can be satisfied with the energy from a portion of the incoming signal. Denote the splitting coefficient at tag $k$ by $\alpha_{k} \in [0, 1]$, which represents the fraction of the incoming signal to be reflected. As such, the remaining $1 - \alpha_{k}$ fraction of the signal energy is used to power the circuit. Where $\xi_{k} = 0$, $\alpha_{k} = 1$.

Denote the channels from the CE to tag $k$, CE to IRS, CE to reader, IRS to tag $k$, reader to IRS and reader to tag $k$ by $\mathbf{h}_{CT_{k}} \in \mathbb{C}^{1 \times L}$, $\mathbf{H}_{CI} \in \mathbb{C}^{N \times L}$, $\mathbf{H}_{CR} \in \mathbb{C}^{M \times L}$, $\mathbf{h}_{T_{k}I}^{H} \in \mathbb{C}^{1 \times N}$, $\mathbf{H}_{RI}^{H} \in \mathbb{C}^{M \times N}$ and $\mathbf{h}_{T_{k}R} \in \mathbb{C}^{M \times 1}$, respectively. Each IRS element $n \in \{1, \ldots, N\}$ reflects the sum of all incident signal paths with unit amplitude gain and a phase shift denoted by $\theta_{n} \in [0, 2\pi)$. The vector containing the phase shifts of all elements is given by $\bm{\theta} = \left[ \theta_{1}, \ldots, \theta_{N} \right]^{T}$. The matrix of reflection coefficient values at the IRS can thus be written as $\mathbf{\Theta} = \mathrm{diag} \left( e^{j \theta_{1}}, \ldots, e^{j \theta_{N}} \right)$.

Linear transmit precoding is assumed at the CE, which transmits signal $s(t)$ to all tags with a single beamforming vector $\mathbf{w}$, such that the transmitted signal can be written as $\mathbf{x}_{C} = \mathbf{w} s(t)$. The transmit power is thus $P = \left\lVert \mathbf{w} \right\rVert^{2}$. The signal received at tag $k$ consists of the combined CE-tag link and is given by
\begin{equation}
y_{T_{k}} = \left( \mathbf{h}_{T_{k}I}^{H} \mathbf{\Theta} \mathbf{H}_{CI} + \mathbf{h}_{CT_{k}} \right) \mathbf{w} s(t).		\label{tagReceivedSignal}
\end{equation}
Note that no noise term exists at the tag, as no signal processing is performed, consistent with e.g., \cite{Wang16}. The part of the signal reflected from the tag is given by $x_{T_{k}} = \sqrt{\alpha_{k}} b(t) y_{T_{k}}$. The remainder of the signal is used to power the circuit, whose energy is given by $(1 - \alpha_{k}) \eta |y_{T_{k}}|^{2}$, where $\eta \in [0, 1]$ is the energy conversion efficiency. For simplicity, $\eta$ is a constant and equal across all tags. Thus, we have the following expression for the circuit constraint:
\begin{equation}
(1 - \alpha_{k}) \eta \left| \left( \mathbf{h}_{T_{k}I}^{H} \mathbf{\Theta} \mathbf{H}_{CI} + \mathbf{h}_{CT_{k}} \right) \mathbf{w} \right|^{2} \geq \xi_{k}.	\label{circuitConstraint}
\end{equation}

The signal received from tag $k$ at the reader consists of the combined tag-reader link:
\begin{multline}
y_{R, k} = \sqrt{\alpha_{k}} b_{k}(t) \mathbf{g}_{k}^{H} \left( \left( \mathbf{H}_{RI}^{H} \mathbf{\Theta} \mathbf{h}_{T_{k}I} + \mathbf{h}_{T_{k}R} \right) \right. \\ \left. \times \left( \mathbf{h}_{T_{k}I}^{H} \mathbf{\Theta} \mathbf{H}_{CI} + \mathbf{h}_{CT_{k}} \right) \mathbf{w} s(t) + \mathbf{n}_{R} \right), \label{readerReceivedSignalFull}
\end{multline}
where $\mathbf{n}_{R} = [n_{R,1}, \ldots, n_{R,M}]^{T}$ is the noise vector at the reader, following $\mathcal{CN}(0, \sigma_{R}^{2} \mathbf{I})$. We assume synchronization errors to be negligible \cite{Wang16}. Note that the full received signal at the reader also contains the direct link terms $\mathbf{H}_{CR} \mathbf{x}_{C}$ and $\mathbf{H}_{RI}^{H} \mathbf{\Theta} \mathbf{H}_{CI} \mathbf{x}_{C}$, which are standalone signal components separate from the tags' signals (i.e., independent of $k$). However, as both are DC terms, they can be removed at the reader before processing, and are hence omitted here.

The instantaneous SNR for tag $k$ at the reader is given by
\begin{multline}
\gamma_{k} = \frac{1}{\sigma_{R}^{2} \left\lVert \mathbf{g}_{k} \right\rVert^{2}} \alpha_{k} |b_{k}(t)|^{2} \left| \mathbf{g}_{k}^{H} \left( \mathbf{H}_{RI}^{H} \mathbf{\Theta} \mathbf{h}_{T_{k}I} + \mathbf{h}_{T_{k}R} \right) \right. \\ \left. \times \left( \mathbf{h}_{T_{k}I}^{H} \mathbf{\Theta} \mathbf{H}_{CI} + \mathbf{h}_{CT_{k}} \right) \mathbf{w} \right|^{2}. 	\label{SNR}
\end{multline}
With fixed subcarrier frequencies $f_{k, 0}$ and $f_{k, 1}$ that are sufficiently distinct, the decoding performance can be improved by maximizing the SNR \cite{Kim14}. Thus, we adopt the SNR as the primary BackCom performance metric for the remainder of this paper.


\section{Problem Formulation}

We study the transmit power minimization problem at the CE subject to each tag's SNR requirement. To do so, we jointly optimize the energy beamforming vector at the CE, the phase shifts at the IRS, the splitting coefficients at each tag and the combining vectors at the reader. We begin by presenting the general IRS-aided multi-tag problem with a multiantenna reader. 

In Section IV, we solve a simplified base case of the general problem with a single semi-passive tag ($\xi = 0$) and a single-antenna reader, which is the most common system setup for bistatic BackCom \cite{Kim14, Fas15, Alev17, Shen16}, and for which closed-form solutions can be derived after problem transformations. The algorithms therein and their associated insights provide the basis for further algorithm development in Section V, where the general problem (whose solutions lack closed-forms and are more difficult to visualize) is studied in greater detail.

Letting $\bm{\alpha} = [\alpha_{1}, \ldots, \alpha_{K}]^{T}$, the general problem can be written as follows:
\begin{subequations}
\begin{align}
\text{(M)}: ~~\min_{\mathbf{w}, \mathbf{\Theta}, \bm{\alpha}, \mathbf{G}} ~~~&\left\lVert \mathbf{w} \right\rVert^{2}		\label{MA} \\
\mathrm{s.t.}~~~~~&\gamma_{k} \geq \gamma_{k, th}, \ \forall k, 	\label{MB} \\
&\left| \left( \mathbf{h}_{T_{k}I}^{H} \mathbf{\Theta} \mathbf{H}_{CI} + \mathbf{h}_{CT_{k}} \right) \mathbf{w} \right|^{2} \nonumber \\
& \qquad \qquad \qquad \geq \frac{\xi_{k}}{(1 - \alpha_{k}) \eta}, \ \forall k, 	\label{MC} \\
&0 \leq \alpha_{k} \leq 1, \ \forall k, 	\label{MD} \\
&0 \leq \theta_{n} < 2 \pi, \ \forall n \in \{1, \ldots, N\},   \label{ME} \\
&\left\lVert \mathbf{g}_{k} \right\rVert^{2} \leq 1, \ \forall k,	\label{MF}
\end{align}
\end{subequations}
where (\ref{MB}) is each tag's SNR constraint, (\ref{MC}) is each tag's circuit power constraint, (\ref{MD}) is the tag splitting coefficient constraint, (\ref{ME}) is the range of phase shifts achievable by each IRS element, and (\ref{MF}) is the constraint on the combining vector for each tag at the reader. To simplify the analysis hereafter, we set $\eta = 1$, $\gamma_{k, th} = \gamma_{th}$ and $\xi_{k} = \xi, \ \forall k$.

The transmit power minimization problem is appealing, as it allows comparisons to be made to the transmit power in a non-IRS-aided system. By fixing a target SNR, one may translate the power reduction to a range increase using the original transmit power. However, Problem (M) is highly nonconvex due to the SNR constraint being a fourth-order function of $\mathbf{\Theta}$, and an optimal solution in terms of all design variables cannot be obtained in a tractable manner. Compared to similar transmit power minimization problems in IRS-aided communication systems, such as that in\cite{Wu19}, Problem (M) has two major differences: 1) the presence of a BackCom device, which is also a passive reflector like the IRS, and poses an additional variable to be optimized (i.e., $\mathbf{\alpha}$); 2) the two-reflection signal path, which renders the problem vastly more complex to solve compared to a single-reflection signal model. In the following sections, we present methods to reduce the complexity of this problem. We consider the non-IRS-assisted system as a benchmark in Section VI to characterize the performance improvements brought about by an IRS.


\section{Base Case: Single Semi-Passive Tag and Single-Antenna Reader}

We begin our study of the general problem with one semi-passive tag and a single-antenna reader. The single-tag setup, as studied in experimental works such as \cite{Fas15, Alev17}, allows us to highlight the characteristics of the solution to the IRS phase shift optimization problem with two reflections of the same signal, both in closed-form (as shown in this section) and visually (as in Section VI) as opposed to the general case, which merits its status as a special case of the general problem. In the following, we propose algorithms to solve the base case problem, to provide the foundation for the general problem in the next section.

For a single-antenna reader, the channels are changed to $\mathbf{h}_{CR}\!\in\!\mathbb{C}^{1 \times L}$, $\mathbf{h}_{RI}^{H}\!\in\!\mathbb{C}^{1 \times N}$ and $h_{TR}\!\in\!\mathbb{C}^{1 \times 1}$. With one semi-passive tag in the system, we can drop the tag indexing  and combiners in Problem~(M), and set the tag splitting coefficient $\alpha$ to $1$. We rewrite the problem as follows:
\begin{subequations}
\begin{align}
\text{(S)}: ~\min_{\mathbf{w}, \mathbf{\Theta}} ~~~&\left\lVert \mathbf{w} \right\rVert^{2}     \label{SA} \\
\mathrm{s.t.}~~~&|b(t)|^{2} \big|\!\left( \mathbf{h}_{RI}^{H} \mathbf{\Theta} \mathbf{h}_{TI} + h_{TR} \right) \nonumber \\ &\quad \times \left( \mathbf{h}_{TI}^{H} \mathbf{\Theta} \mathbf{H}_{CI} + \mathbf{h}_{CT} \right) \mathbf{w} \big|^{2} \geq \gamma_{th} \sigma_{R}^{2}, 	\label{SB} \\
&0 \leq \theta_{n} < 2 \pi, \ \forall n \in \{1, \ldots, N\}. \label{SC}
\end{align}
\end{subequations}
Here, the phase shift matrix $\mathbf{\Theta}$ needs to be designed to balance the channel gains between the combined CE-tag and combined tag-reader links. For the single-tag case, it is well-known that optimal beamforming can be achieved by using maximum ratio transmission (MRT), given by\footnote{Note that when $\xi > 0$, MRT is not necessarily optimal, as constraint (\ref{MC}) also becomes active. A solution for $\mathbf{w}^{*}$ can then be found using the approaches proposed in Section V.}
\begin{equation}
\mathbf{w}^{*} = \sqrt{P} \frac{\left[ \left( \mathbf{h}_{RI}^{H} \mathbf{\Theta} \mathbf{h}_{TI} + h_{TR} \right) \left( \mathbf{h}_{TI}^{H} \mathbf{\Theta} \mathbf{H}_{CI} + \mathbf{h}_{CT} \right) \right]^{H}}{\left\lVert  \left( \mathbf{h}_{RI}^{H} \mathbf{\Theta} \mathbf{h}_{TI} + h_{TR} \right) \left( \mathbf{h}_{TI}^{H} \mathbf{\Theta} \mathbf{H}_{CI} + \mathbf{h}_{CT} \right) \right\rVert}. 		\label{optimalBV}
\end{equation}
Substituting $\mathbf{w}^{*}$ into Problem (S), we rewrite it in power minimization form:
\begin{subequations}
\begin{align}
\text{(S1)}: ~~\min_{P, \mathbf{\Theta}} ~~~&P 	   \label{S1A} \\
\mathrm{s.t.}~~~~&P |b(t)|^{2} \big\lVert \left( \mathbf{h}_{RI}^{H} \mathbf{\Theta} \mathbf{h}_{TI} + h_{TR} \right) \nonumber \\ &\quad \times \left( \mathbf{h}_{TI}^{H} \mathbf{\Theta} \mathbf{H}_{CI} + \mathbf{h}_{CT} \right) \big\rVert^{2} \geq \gamma_{th} \sigma_{R}^{2}, 	\label{S1B} \\
& \text{(\ref{SC})}. \nonumber
\end{align}
\end{subequations}
By inspection, the optimal transmit power is the minimum value that satisfies (\ref{S1B}):
\begin{equation}
P^{*} = \frac{\gamma_{th} \sigma_{R}^{2}}{|b(t)|^{2} \left\lVert \left( \mathbf{h}_{RI}^{H} \mathbf{\Theta} \mathbf{h}_{TI} + h_{TR} \right) \left( \mathbf{h}_{TI}^{H} \mathbf{\Theta} \mathbf{H}_{CI} + \mathbf{h}_{CT} \right) \right\rVert^{2}}. \label{optimalPower}
\end{equation}
To obtain $P^{*}$, we can directly maximize the denominator over $\mathbf{\Theta}$, with the problem as follows:
\begin{align}
\text{(S2)}: ~~\max_{\mathbf{\Theta}} ~~~&\left\lVert \left( \mathbf{h}_{RI}^{H} \mathbf{\Theta} \mathbf{h}_{TI} + h_{TR} \right) \left( \mathbf{h}_{TI}^{H} \mathbf{\Theta} \mathbf{H}_{CI} + \mathbf{h}_{CT} \right) \right\rVert^{2}, \label{S2}  \\
\mathrm{s.t.}~~~~&\text{(\ref{SC})}.	\nonumber
\end{align}

\subsection{Minorization Maximization Algorithm}

To solve Problem (S2), we may split (\ref{S2}) into a squared norm and a scalar term, corresponding to the two bracketed terms. The two terms can be expanded as
\begin{multline}
\left\lVert \mathbf{h}_{TI}^{H} \mathbf{\Theta} \mathbf{H}_{CI} + \mathbf{h}_{CT} \right\rVert^{2} = \mathbf{v}^{H} \mathbf{\Phi}_{CIT} \mathbf{\Phi}_{CIT}^{H} \mathbf{v} \\ + \mathbf{v}^{H} \mathbf{\Phi}_{CIT} \mathbf{h}_{CT}^{H} + \mathbf{h}_{CT} \mathbf{\Phi}_{CIT}^{H} \mathbf{v} + \left\lVert \mathbf{h}_{CT} \right\rVert^{2},		\label{QCQPStandard}
\end{multline}
\begin{multline}
|\mathbf{h}_{RI}^{H} \mathbf{\Theta} \mathbf{h}_{TI} + h_{TR}|^{2} = \mathbf{v}^{H} \mathbf{\Phi}_{TIR} \mathbf{\Phi}_{TIR}^{H} \mathbf{v} \\ + \mathbf{v}^{H} \mathbf{\Phi}_{TIR} h_{TR} + h_{TR}^{H} \mathbf{\Phi}_{TIR}^{H} \mathbf{v} + |h_{TR}|^{2}, \label{QCQP2}
\end{multline}
where $\mathbf{\Phi}_{CIT} = \mathrm{diag}(\mathbf{h}_{TI}^{H}) \mathbf{H}_{CI}$, $\mathbf{\Phi}_{TIR} = \mathrm{diag}(\mathbf{h}_{RI}^{H}) \mathbf{h}_{IT}$, and $\mathbf{v} = \left[ e^{j \theta_{1}}, \ldots, e^{j \theta_{N}} \right]^{H}$, where $|v_{n}|^{2} = 1, \ \forall n$. These two equations can be rewritten in matrix form as
\begin{align}
\left\lVert \mathbf{h}_{TI}^{H} \mathbf{\Theta} \mathbf{H}_{CI} + \mathbf{h}_{CT} \right\rVert^{2} &= \bar{\mathbf{v}}^{H} \mathbf{R} \bar{\mathbf{v}} + \left\lVert \mathbf{h}_{CT} \right\rVert^{2},	\label{expandedQCQP} \\
|\mathbf{h}_{RI}^{H} \mathbf{\Theta} \mathbf{h}_{TI} + h_{TR}|^{2} &= \mathbf{\bar{v}}^{H} \mathbf{S} \mathbf{\bar{v}} + |h_{TR}|^{2},
\end{align}
with
\begin{align}
\mathbf{R} &= 
\begin{bmatrix}
\mathbf{\Phi}_{CIT} \mathbf{\Phi}_{CIT}^{H} & \mathbf{\Phi}_{CIT} \mathbf{h}_{CT}^{H} \\
\mathbf{h}_{CT} \mathbf{\Phi}_{CIT}^{H} & 0
\end{bmatrix}, \nonumber \\
\mathbf{S} &= 
\begin{bmatrix}
\mathbf{\Phi}_{TIR} \mathbf{\Phi}_{TIR}^{H} & \mathbf{\Phi}_{TIR} h_{TR}^{H} \\
h_{TR} \mathbf{\Phi}_{TIR}^{H} & 0
\end{bmatrix},
\hspace{10mm} \mathbf{\bar{v}} = 
\begin{bmatrix}
\mathbf{v} \\
1
\end{bmatrix}.  \label{R}
\end{align}
As a result, the product of (\ref{expandedQCQP}) and (\ref{QCQP2}) can be written as
\begin{equation}
F(\bar{\mathbf{v}}) = \bar{\mathbf{v}}^{H} \mathbf{S} \bar{\mathbf{v}} \bar{\mathbf{v}}^{H} \mathbf{R} \bar{\mathbf{v}} + c_{1} \bar{\mathbf{v}}^{H} \mathbf{S} \bar{\mathbf{v}} + c_{2} \bar{\mathbf{v}}^{H} \mathbf{R} \bar{\mathbf{v}} + c_{1} c_{2}, \label{revisedOF}
\end{equation}
with $c_{1} = \lVert \mathbf{h}_{CT} \rVert^{2}$ and $c_{2} = |h_{TR}|^{2}$. Equation (\ref{revisedOF}) is a quartic polynomial in $\bar{\mathbf{v}}$. Normally, IRS phase shift optimization problems involve optimization over a quadratic polynomial such as (\ref{expandedQCQP}), which permits the use of the identity $\bar{\mathbf{v}}^{H} \mathbf{R} \bar{\mathbf{v}} = \mathrm{tr}(\mathbf{R} \bar{\mathbf{v}} \bar{\mathbf{v}}^{H})$. If we let $\mathbf{V} = \mathbf{\bar{v}} \mathbf{\bar{v}}^{H}$, then the objective function becomes a function of $\mathbf{V}$ (i.e., $\mathrm{tr}(\mathbf{R} \mathbf{V})$), which is rank-one. Here, we cannot invoke this identity, as the first resulting trace term, $\mathrm{tr}(\mathbf{S} \mathbf{V} \mathbf{R} \mathbf{V})$, is generally nonconvex, as $\mathbf{R}$ and $\mathbf{S}$ are generally not positive semidefinite. It has also been noted in the literature that the optimization (minimization, in the case of \cite{Luo10}) of multivariate polynomials of degree $4$ and above is NP-hard, meaning that a closed-form, optimal solution is generally not available. We address this challenging issue by using the MM algorithm.

To solve Problem (S2), in each MM iteration, we first find a minorizer to $F(\bar{\mathbf{v}})$, and solve the maximization problem with the minorizer as the objective. We note that in \cite[Lemma 12]{Sun17}, a convex \textit{majorizing} function was derived for functions $f(\mathbf{x}): \mathbb{R}^{N} \rightarrow \mathbb{R}$. As our objective is $f(\mathbf{x}): \mathbb{C}^{N} \rightarrow \mathbb{R}$, we can use similar logic and complex calculus to obtain a \textit{minorizer} with bounded curvature by taking the first-order Taylor expansion plus a \textit{negative} squared error term:
\begin{equation}
f(\mathbf{x}) \geq f(\mathbf{x}_{0}) + \mathrm{Re} \left\{ \nabla f(\mathbf{x}_{0})^{H} (\mathbf{x} - \mathbf{x}_{0}) \right\} - \frac{\ell}{2} \left\lVert \mathbf{x} - \mathbf{x}_{0} \right\rVert^{2}, 	\label{minoriser}
\end{equation}
where $\mathbf{x}_{0} \in \mathbb{C}^{N}$ is the point of intersection between $f(\mathbf{x})$ and the minorizer, $\nabla$ is the gradient operator, and $\ell$ is the Lipschitz constant (i.e., maximum curvature of $f(\mathbf{x})$). Following (\ref{minoriser}), the minorizer to (\ref{revisedOF}) is
\begin{align}
F(\bar{\mathbf{v}}) &\geq \bar{\mathbf{v}}_{0}^{H} \mathbf{S} \bar{\mathbf{v}}_{0} \bar{\mathbf{v}}_{0}^{H} \mathbf{R} \bar{\mathbf{v}}_{0} + c_{1} \bar{\mathbf{v}}_{0}^{H} \mathbf{S} \bar{\mathbf{v}}_{0} + c_{2} \bar{\mathbf{v}}_{0}^{H} \mathbf{R} \bar{\mathbf{v}}_{0} \nonumber \\
		& \qquad + c_{1} c_{2} + \bar{\mathbf{v}}_{0}^{H} \mathbf{T} (\bar{\mathbf{v}} - \bar{\mathbf{v}}_{0}) + (\bar{\mathbf{v}} - \bar{\mathbf{v}}_{0})^{H} \mathbf{T} \bar{\mathbf{v}}_{0}  \nonumber \\
		& \qquad - \frac{\ell}{2}(\bar{\mathbf{v}}^{H} \bar{\mathbf{v}} - \bar{\mathbf{v}}^{H} \bar{\mathbf{v}}_{0} - \bar{\mathbf{v}}_{0}^{H} \bar{\mathbf{v}} + \left\lVert \bar{\mathbf{v}}_{0} \right\rVert^{2}) \nonumber \\
		&= -\frac{\ell}{2}(\bar{\mathbf{v}}^{H} \bar{\mathbf{v}} - \bar{\mathbf{v}}^{H} \bar{\mathbf{v}}_{0} - \bar{\mathbf{v}}_{0}^{H} \bar{\mathbf{v}} + \left\lVert \bar{\mathbf{v}}_{0} \right\rVert^{2}) \nonumber \\
		& \qquad + \bar{\mathbf{v}}_{0}^{H} \mathbf{T} \bar{\mathbf{v}} + \bar{\mathbf{v}}^{H} \mathbf{T} \bar{\mathbf{v}}_{0} + c \nonumber \\
		&= -\frac{\ell}{2} \left( \bar{\mathbf{v}}^{H} \mathbf{I} \bar{\mathbf{v}} + \bar{\mathbf{v}}^{H} \left( -\frac{2}{\ell} \mathbf{T} \bar{\mathbf{v}}_{0} - \mathbf{I} \bar{\mathbf{v}}_{0} \right) \right. \nonumber \\
		& \left. \qquad + \left( -\frac{2}{\ell} \mathbf{T} \bar{\mathbf{v}}_{0} - \mathbf{I} \bar{\mathbf{v}}_{0} \right)^{H} \mathbf{\bar{v}} \right) + c, 	\label{quadraticApprox}
\end{align}
with $\mathbf{T} = \mathbf{R} \bar{\mathbf{v}}_{0} \bar{\mathbf{v}}_{0}^{H} \mathbf{S} + \mathbf{S} \bar{\mathbf{v}}_{0} \bar{\mathbf{v}}_{0}^{H} \mathbf{R} + c_{2} \mathbf{R} + c_{1} \mathbf{S}$ being a Hermitian matrix obtained from the derivative of $F(\bar{\mathbf{v}}_{0})$, and $c$ denoting the cumulative sum of all constant terms and terms involving only $\bar{\mathbf{v}}_{0}$. Equation (\ref{quadraticApprox}) is also of quadratic form, and can be rewritten as $\bar{\bar{\mathbf{v}}}^{H} \mathbf{U} \bar{\bar{\mathbf{v}}}$, where
\begin{equation}
\mathbf{U} = 
-\begin{bmatrix}
\mathbf{I} & -\frac{2}{\ell} \mathbf{T} \bar{\mathbf{v}}_{0} - \mathbf{I} \bar{\mathbf{v}}_{0} \\
(-\frac{2}{\ell} \mathbf{T} \bar{\mathbf{v}}_{0} - \mathbf{I} \bar{\mathbf{v}}_{0})^{H} & 0
\end{bmatrix},
\hspace{2mm} \bar{\bar{\mathbf{v}}} = 
\begin{bmatrix}
\bar{\mathbf{v}} \\
1
\end{bmatrix}.  \label{U}
\end{equation}

At this point, one approach is to let $\bar{\bar{\mathbf{V}}} = \bar{\bar{\mathbf{v}}} \bar{\bar{\mathbf{v}}}^{H}$, where $\bar{\bar{\mathbf{V}}} \succeq 0$ and $\mathrm{rank}(\bar{\bar{\mathbf{V}}}) = 1$. This gives us the minorizer $\mathrm{tr}(\mathbf{U} \bar{\bar{\mathbf{V}}}) + c$ to be maximized with respect to $\bar{\bar{\mathbf{V}}}$ in each MM iteration. Relaxing the rank constraint, we obtain a semidefinite program (SDP) that can be optimally solved using CVX \cite{cvx}. The best resulting $\bar{\bar{\mathbf{v}}}$ from Gaussian randomization \cite{gaussrand} can then be converted to a rank-one $\bar{\bar{\mathbf{V}}}$. Nonetheless, the SDR approach by itself incurs a worst-case complexity on the order of $O(I(N+2)^{4.5}) \simeq O(IN^{4.5})$ \cite{Luo10a}, where $I$ is the number of MM iterations. This complexity may be avoided by taking a further minorizer to $\bar{\bar{\mathbf{v}}}^{H} \mathbf{U} \bar{\bar{\mathbf{v}}}$ as in \cite{Pan19}:
\begin{equation}
\bar{\bar{\mathbf{v}}}^{H} \mathbf{U} \bar{\bar{\mathbf{v}}} \geq \bar{\bar{\mathbf{v}}}^{H} \mathbf{X} \bar{\bar{\mathbf{v}}} + 2\mathrm{Re}\{\bar{\bar{\mathbf{v}}}^{H} (\mathbf{U} - \mathbf{X}) \bar{\bar{\mathbf{v}}}_{0}\} + \bar{\bar{\mathbf{v}}}_{0}^{H} (\mathbf{X} - \mathbf{U}) \bar{\bar{\mathbf{v}}}_{0},		\label{secondMinoriser}
\end{equation}
with $\mathbf{X} = \lambda^{-} \mathbf{I}$, $\lambda^{-}$ being the minimum eigenvalue of $\mathbf{U}$, and $\bar{\bar{\mathbf{v}}}_{0}$ as before. The new objective is to maximize the second minorizer in every MM iteration, with respect to $\bar{\bar{\mathbf{v}}}$. As the first and last terms of (\ref{secondMinoriser}) are both constants, each MM subproblem reduces to
\begin{subequations}
\begin{align}
\text{(S3)}: ~~\max_{\bar{\bar{\mathbf{v}}}} ~~~&2\mathrm{Re}\{\bar{\bar{\mathbf{v}}}^{H} (\mathbf{U} - \mathbf{X}) \bar{\bar{\mathbf{v}}}_{0}\}, 	\label{S3MA}	\\
\mathrm{s.t.}~~~~&|\bar{\bar{\mathbf{v}}}_{n, n}| = 1, \ \forall n \in \{1, \ldots, N+2\}, \label{S3MB}  \\
&\bar{\bar{\mathbf{v}}}_{j, j} = 1, \ j = \{N+1, N+2\}.	\label{S3MC}
\end{align}
\end{subequations}
Problem (S3) has a closed-form solution, given by $\bar{\bar{\mathbf{v}}}^{*} = e^{j \ \mathrm{arg}((\mathbf{U} - \mathbf{X}) \bar{\bar{\mathbf{v}}}_{0})}$.

Note that $F(\bar{\mathbf{v}})$ is bounded above, as both $\mathbf{R}$ and $\mathbf{S}$ are constant matrices, and $\left\lVert \bar{\mathbf{v}} \right\rVert^{2}\!=\!N$ is a finite constant. Both minorizers can be readily shown to satisfy the conditions required for MM convergence  outlined in \cite{Pan19, Sun17} with respect to their objective functions. In addition, the second minorizer also acts as minorizer to the original objective function in (\ref{revisedOF}), and returns an optimal solution in terms of $\bar{\bar{\mathbf{v}}}$ in each MM iteration. Thus, the sequence of solutions from each iteration of Problem (S3) will monotonically increase and converge to at least a local optimum of (\ref{revisedOF}). \textbf{Algorithm 1} summarizes the process.

\begin{algorithm}
	\caption{MM Algorithm with Nested SDR}
	\begin{algorithmic}[1]
		\STATE \textbf{Initialize:} Random IRS phase shifts $\bm{\theta}$; set iteration number $i = 1$.
		\STATE Obtain $\bar{\mathbf{v}}$ from $\bm{\theta}$ and set $\bar{\mathbf{v}}_{0}^{(i)} = \bar{\mathbf{v}}$.
		\WHILE{the rate of change in objective function (\ref{revisedOF}) is above a threshold $\varepsilon > 0$}
			\STATE Construct $\mathbf{U}$ from $\bar{\mathbf{v}}_{0}^{(i)}$ and $\mathbf{T}$.
			\STATE Set $\bar{\mathbf{v}}_{0}^{(i+1)} \leftarrow \bar{\bar{\mathbf{v}}}^{*} = e^{j \ \mathrm{arg}((\mathbf{U} - \mathbf{X}) \bar{\bar{\mathbf{v}}}_{0})}$ for next iteration after removing the last element.
			\STATE Update iteration number $i \leftarrow i + 1$.
		\ENDWHILE
		\STATE \textbf{Return:} Optimized phase shift vector $\mathbf{v}^{*}$ by dropping the last element of $\bar{\mathbf{v}}_{0}$ at convergence.
	\end{algorithmic}
\end{algorithm}

\subsection{Successive Refinement Algorithm}

We present a successive refinement (SR)-based phase shift optimization algorithm, where in each iteration, the $N$ phase shifts are optimized sequentially starting from the first, while holding the others constant. Denoting the current reflection coefficient to be optimized as $s_{n} = e^{j \theta_{n}}$, the objective function to be maximized with respect to each phase shift is given by
\begin{align}
F(s_{n}) &= \left| \mathbf{h}_{RI}^{H} \mathbf{\Theta} \mathbf{h}_{TI} \mathbf{h}_{TI}^{H} \mathbf{\Theta} \mathbf{H}_{CI} \mathbf{w} + \mathbf{h}_{RI}^{H} \mathbf{\Theta} \mathbf{h}_{TI} \mathbf{h}_{CT} \mathbf{w} \right. \nonumber \\
		& \left. + h_{TR} \mathbf{h}_{TI}^{H} \mathbf{\Theta} \mathbf{H}_{CI} \mathbf{w} + h_{TR} \mathbf{h}_{CT} \mathbf{w} \right|^{2} \nonumber \\
		&=\!\left| \left[ [\mathbf{h}_{RI}^{H}]_{n} s_{n} [\mathbf{h}_{TI}]_{n}\!+\!\sum_{j \neq n}^{N} [\mathbf{h}_{RI}^{H}]_{j} s_{j} [\mathbf{h}_{TI}]_{j} \right] \right. \nonumber \\
		& \left.\!\times \left[ [\mathbf{h}_{TI}^{H}]_{n} s_{n} [\mathbf{H}_{CI} \mathbf{w}]_{n}\!+\!\sum_{j \neq n}^{N} [\mathbf{h}_{TI}^{H}]_{j} s_{j} [\mathbf{H}_{CI} \mathbf{w}]_{j} \right] \right. \nonumber \\
		& \left.\!+ \left[ [\mathbf{h}_{RI}^{H}]_{n} s_{n} [\mathbf{h}_{TI}]_{n} \mathbf{h}_{CT} \mathbf{w}\!+\!\sum_{j \neq n}^{N} [\mathbf{h}_{RI}^{H}]_{j} s_{j} [\mathbf{h}_{TI}]_{j} \mathbf{h}_{CT} \mathbf{w} \right] \right. \nonumber \\
		& \left.\!+ \left[ [h_{TR} \mathbf{h}_{TI}^{H}]_{n} s_{n} [\mathbf{H}_{CI} \mathbf{w}]_{n}\!+\!\sum_{j \neq n}^{N} [h_{TR} \mathbf{h}_{TI}^{H}]_{j} s_{j} [\mathbf{H}_{CI} \mathbf{w}]_{j} \right] \right. \nonumber \\
		& \left. + h_{TR} \mathbf{h}_{CT} \mathbf{w} \right|^{2}.  \label{expand}
\end{align}
The exact solution $\theta_{n}^{*}$ may be found by linear search. Alternatively, we can also operate further on (\ref{expand}), which can be rewritten as
\begin{equation}
F(s_{n}) = \left| h_{1} s_{n}^{2} + (h_{2} + h_{3}) s_{n} + h_{4} \right|^{2}, 	\label{expand2}
\end{equation}
where $h_{1}$ is the product of the $s_{n}$ terms in the first two brackets of (\ref{expand}); $h_{2}$ and $h_{3}$ are the non-summation terms in the second and third brackets, respectively; and $h_{4}$ is the sum of all remaining constants. Let $h_{\Sigma} = h_{2} + h_{3}$. When expanded, (\ref{expand2}) can be rewritten as
\begin{multline}
F(\theta_{n}) = 2 \mathrm{Re}\left\{ a_{1 \Sigma} e^{j \theta_{n}} e^{j \theta_{1 \Sigma}} \right\} + 2 \mathrm{Re}\left\{ a_{\Sigma 4} e^{j \theta_{n}} e^{j \theta_{\Sigma 4}} \right\} \\ + 2 \mathrm{Re}\left\{ a_{1 4} e^{j 2 \theta_{n}} e^{j \theta_{14}} \right\} + \kappa,	\label{Fsj}
\end{multline}
where $a_{pq}$ and $\theta_{pq}$ denote the magnitude and phase of the composite term $h_{p} h_{q}^{H}, p, q \in \{1, \Sigma, 4\}$, and $\kappa = |h_{1}|^{2} + |h_{\Sigma}|^{2} + |h_{4}|^{2}$. With the reference path loss magnitude incorporated into the channels, one may note that $h_{4}$ is larger than both $h_{\Sigma}$ and $h_{1}$ by several orders of magnitude, since $h_{\Sigma}$ and $h_{1}$ are comprised of three and four channel terms, respectively, compared to two for $h_{4}$. Hence, the product magnitude of $h_{\Sigma}$ and $h_{4}$ dominates the objective function (out of the terms involving $\theta_{n}$). As such, the approximate derivative of $F(\theta_{n})$ can be given by
\begin{equation}
F_{apx}^{'}(\theta_{n}) = -2 \sin \left( \theta_{n} + \theta_{\Sigma 4} \right), \label{apx}
\end{equation}
which leads to an approximate closed-form solution of $\theta_{n}^{*} \approx \mathrm{mod}(2 \pi - \theta_{\Sigma 4}, 2 \pi)$.

The successive refinement of individual phase shifts is guaranteed to converge, as $F(s_{n})$ is nondecreasing due to the coupled nature of all phase shifts. While the converged solution may not be a local optimum, our numerical results in Section VI-A suggest close agreement between this algorithm and the MM-based algorithm in the previous section, whose solution has the optimality guarantee. Upon convergence of $F(s_{n})$, $\mathbf{w}$ is computed using (\ref{optimalBV}). This alternating process repeats until convergence of the objective function $\left\lVert \mathbf{w} \right\rVert^{2}$. 

The complexity of \textbf{Algorithm 1} in the previous subsection is on the order of $O(I_{MM} N^{2})$, and that of the SR approach in this subsection is $O(I_{SR} N^{2})$, where $I_{MM}$ and $I_{SR}$ are the number of MM sub-iterations and SR cycles through all phase shifts, respectively.


\section{General Case: Multiple Tags and Multiantenna Reader}

The bistatic BackCom system with a multiantenna reader and multiple (possibly passive) tags encompasses all practical system configurations. However, as the IRS needs to be optimized to cater to multiple tags each with their own twice-reflected signal, the optimization problem becomes noticeably more complex. Given the coupling of $\mathbf{w}$, $\mathbf{\Theta}$, $\bm{\alpha}$ and $\mathbf{g}_{k}$ in (\ref{MB}), we use an AO approach to solve Problem (M), which iteratively optimizes one variable while holding the others constant. To optimize $\mathbf{\Theta}$ over multiple tags, we directly draw upon, and upscale the MM algorithm in Section IV-A as an intuitive approach. We also propose a low-complexity algorithm based on the successive refinement algorithm in Section IV-B.

\subsection{Transmit Beamforming Vector Optimization}

First, we minimize the squared norm of the energy beamforming vector $\mathbf{w}$:
\begin{subequations}
\begin{align}
\text{(M1)}: ~~\min_{\mathbf{w}} ~&\left\lVert \mathbf{w} \right\rVert^{2} 		\label{M1A}\\
\mathrm{s.t.}~~&\big| \mathbf{g}_{k}^{H}\!\left( \mathbf{H}_{RI}^{H} \mathbf{\Theta} \mathbf{h}_{T_{k}I}\!+\!\mathbf{h}_{T_{k}R} \right) \nonumber \\
& \quad \times \left( \mathbf{h}_{T_{k}I}^{H} \mathbf{\Theta} \mathbf{H}_{CI}\!+\!\mathbf{h}_{CT_{k}} \right)\!\mathbf{w} \big|^{2} \geq \frac{\alpha_{k} |b_{k}(t)|^{2}}{\gamma_{th} \sigma_{R}^{2}}, \ \forall k, \label{M1B} \\
&\left| \left( \mathbf{h}_{T_{k}I} \mathbf{\Theta} \mathbf{H}_{CI} + \mathbf{h}_{CT_{k}} \right) \mathbf{w} \right|^{2} \geq \frac{\xi}{1 - \alpha_{k}}, \ \forall k. \label{M1C}
\end{align}
\end{subequations}
We let $\mathbf{W} = \mathbf{w} \mathbf{w}^{H}$, where $\mathbf{W} \succeq 0$ and $\mathrm{rank}(\mathbf{W}) = 1$. In addition, where appropriate, we make the substitutions $\mathbf{h}_{k, 1}(\mathbf{\Theta}) = \mathbf{h}_{T_{k}I}^{H} \mathbf{\Theta} \mathbf{H}_{CI} + \mathbf{h}_{CT_{k}}$ and $\mathbf{h}_{k, 2}(\mathbf{\Theta}) = \mathbf{H}_{RI}^{H} \mathbf{\Theta} \mathbf{h}_{T_{k}I} + \mathbf{h}_{T_{k}R}$. As a result, the objective function (\ref{M1A}) is modified to $\mathrm{tr}(\mathbf{W})$; and the LHS of (\ref{M1B}) can be rewritten as $|\mathbf{g}_{k}^{H} \mathbf{h}_{k, 2}(\mathbf{\Theta})|^{2} \mathrm{tr}(\mathbf{W} \mathbf{H}_{k, 1}(\mathbf{\Theta}))$, with $\mathbf{H}_{k, 1}(\mathbf{\Theta}) = \mathbf{h}_{k, 1}(\mathbf{\Theta})^{H} \mathbf{h}_{k, 1}(\mathbf{\Theta}) \in \mathbb{C}^{L \times L}$. The LHS of (\ref{M1C}) can be rewritten as $(1 - \alpha_{k}) \mathrm{tr}(\mathbf{W} \mathbf{H}_{k, 1}(\mathbf{\Theta}))$. Thus, dropping the nonconvex rank constraint on $\mathbf{W}$, Problem (M1) can be transformed to the following problem:
\begin{subequations}
\begin{align}
\text{(M1.1)}: ~~\min_{\mathbf{W}} ~&\mathrm{tr}(\mathbf{W}) \label{M11A} \\
\mathrm{s.t.}~~&\big| \mathbf{g}_{k}^{H} \mathbf{h}_{k, 2}(\mathbf{\Theta}) \big|^{2} \mathrm{tr}(\mathbf{W} \mathbf{H}_{k, 1}(\mathbf{\Theta})) \nonumber \\
& \qquad \qquad \qquad \geq \frac{\gamma_{th} \sigma_{R}^{2}}{\alpha_{k} |b_{k}(t)|^{2}}, \ \forall k, 	\label{M11B} \\
&\mathrm{tr}(\mathbf{W} \mathbf{H}_{k, 1}(\mathbf{\Theta})) \geq \frac{\xi}{1 - \alpha_{k}}, \ \forall k, \label{M11C} \\
&\mathbf{W} \succeq 0.   \label{M11D}
\end{align}
\end{subequations}
Problem (M1.1) is an SDP, and can be solved optimally using CVX. After obtaining a candidate solution $\mathbf{W}_{SDR}$, Gaussian randomization can be performed to obtain the best beamforming vector $\mathbf{w}_{SDR}^{*}$ to obtain a rank-one $\mathbf{W}_{SDR}$. Each beamforming vector generated by randomization is appropriately scaled to satisfy the most violated constraint for all tags out of (\ref{M11B})-(\ref{M11C}) with equality \cite{gaussrand}. The best solution is thus the post-scaling vector with the smallest norm.

\subsection{IRS Phase Shift Optimization}

As the objective function (\ref{MA}) in Problem (M) only depends on $\mathbf{w}$, the optimization over $\mathbf{\Theta}$ takes the form of a feasibility problem, which can be written as
\begin{subequations}
\begin{align}
\text{(M2)}: ~~\mathrm{find} ~~&\mathbf{\Theta} 	\label{M2A} \\
\mathrm{s.t.}~~&\left| \mathbf{g}_{k}^{H} \mathbf{h}_{k, 2}(\mathbf{\Theta}) \mathbf{h}_{k, 1}(\mathbf{\Theta}) \mathbf{w} \right|^{2} \geq \frac{\gamma_{th} \sigma_{R}^{2}}{\alpha_{k} |b_{k}(t)|^{2}}, \ \forall k, \label{M2B} \\
&\left| \mathbf{h}_{k, 1}(\mathbf{\Theta}) \mathbf{w} \right|^{2} \geq \frac{\xi}{1 - \alpha_{k}}, \ \forall k, \label{M2C} \\
&0 \leq \theta_{n} < 2 \pi, \	\forall n. \label{M2D}
\end{align}
\end{subequations}
The quartic function of $\mathbf{\Theta}$ in (\ref{M2B}) makes this problem highly nonconvex. Similar to the base case, we present two methods to finding a feasible $\mathbf{\Theta}$. The first involves a series of transformations on Problem (M2) to a more tractable, approximating problem, on which we can apply an MM-based approach to reach a desirable solution to the overall problem. The second is an extension of the successive phase shift optimization algorithm presented in Section IV-B.

\textit{1) MM-based Algorithm:} In this first method, we approximate each quartic constraint with a simpler minorizing constraint in each iteration. First, we recast the squared magnitude expression in (\ref{M2B}) using the same $\mathbf{v}$ and $\bar{\mathbf{v}}$ introduced in (\ref{QCQPStandard}) and (\ref{R}), respectively, resulting in 
\begin{equation}
\bar{\mathbf{v}}^{H} \mathbf{A}_{k} \bar{\mathbf{v}} \bar{\mathbf{v}}^{H} \mathbf{C}_{k} \bar{\mathbf{v}} + |d_{k}|^{2} \bar{\mathbf{v}}^{H} \mathbf{A}_{k} \bar{\mathbf{v}} + |b_{k}|^{2} \bar{\mathbf{v}}^{H} \mathbf{C}_{k} \bar{\mathbf{v}} + |b_{k}|^{2} |d_{k}|^{2},		\label{modifiedM2B}
\end{equation}
where $\mathbf{a}_{k} = \text{diag}(\mathbf{g}_{k}^{H} \mathbf{H}_{RI}^{H}) \mathbf{h}_{T_{k}I}$, $b_{k} = \mathbf{g}_{k}^{h} \mathbf{h}_{T_{k}R}$, $\mathbf{c}_{k} = \text{diag}(\mathbf{h}_{T_{k}I}^{H}) \mathbf{H}_{CI} \mathbf{w}$, $d_{k} = \mathbf{h}_{CT_{k}} \mathbf{w}$, and
\begin{equation}
\mathbf{A}_{k} = 
\begin{bmatrix}
\mathbf{a}_{k} \mathbf{a}_{k}^{H} & b_{k}^{H} \mathbf{a}_{k} \\
b_{k} \mathbf{a}_{k}^{H} & 0
\end{bmatrix},
\hspace{10mm} \mathbf{C}_{k} = 
\begin{bmatrix}
\mathbf{c}_{k} \mathbf{c}_{k}^{H} & d_{k}^{H} \mathbf{c}_{k}  \\
d_{k} \mathbf{c}_{k}^{H} & 0
\end{bmatrix}.  \label{AC}
\end{equation}
The circuit constraint in (\ref{M2C}) can be similarly recast as
\begin{equation}
(1 - \alpha_{k}) (\bar{\mathbf{v}}^{H} \mathbf{C}_{k} \bar{\mathbf{v}} + |d_{k}|^{2}) \geq \xi. 	\label{modifiedM2C}
\end{equation}
Then, we formulate convex versions of both (\ref{modifiedM2B}) and (\ref{modifiedM2C}). For each tag, the SNR constraint is approximated by a convex minorizer in each iteration similar to the MM procedure from Section IV, which is a stricter version of the original constraint. Holding $\alpha_{k}$ and $b_{k}(t)$ constant, (\ref{modifiedM2B}) can be lower-bounded by $\mathrm{tr}( \mathbf{U}_{k} \bar{\bar{\mathbf{V}}} )$, with
\begin{equation}
\mathbf{U}_{k} = 
-\begin{bmatrix}
\mathbf{I} & -\frac{2}{\ell} \mathbf{T}_{k} \bar{\mathbf{v}}_{0} - \mathbf{I} \bar{\mathbf{v}}_{0} \\
(-\frac{2}{\ell} \mathbf{T}_{k} \bar{\mathbf{v}}_{0} - \mathbf{I} \bar{\mathbf{v}}_{0})^{H} & 0
\end{bmatrix},	\label{U_k}
\end{equation}
where $\bar{\bar{\mathbf{V}}}$ was defined in Problem (S3),
$\mathbf{T}_{k} = \mathbf{A}_{k} \bar{\mathbf{v}}_{0} \bar{\mathbf{v}}_{0}^{H} \mathbf{C}_{k} + \mathbf{C}_{k} \bar{\mathbf{v}}_{0} \bar{\mathbf{v}}_{0}^{H} \mathbf{A}_{k} + |b_{k}|^{2} \mathbf{C}_{k} + |d_{k}|^{2} \mathbf{A}_{k}$, and $\bar{\mathbf{v}}_{0} \in \mathbb{C}^{N+1}$ is an initialization vector with $1$ as its last element. The new constraint is
\begin{equation}
\alpha_{k} |b_{k}(t)|^{2} \left( \mathrm{tr}(\mathbf{U}_{k} \bar{\bar{\mathbf{V}}}) + |b_{k}|^{2} |d_{k}|^{2} \right) \geq \gamma_{th} \sigma_{R}^{2}. 	\label{modifiedQ3B}
\end{equation}
The circuit constraint in (\ref{modifiedM2C}) can be recast into a convex form by defining a new matrix $\bar{\mathbf{C}}_{k}$, whose dimensions match that of $\mathbf{U}_{k}$ by adding an extra zero row and column:
\begin{equation}
(1 - \alpha_{k}) \left( \mathrm{tr}(\bar{\mathbf{C}}_{k} \bar{\bar{\mathbf{V}}}) + |d_{k}|^{2}\right) \geq \xi.
\label{modifiedQ3C}
\end{equation}
With these transformations, we obtain a feasibility problem in $\bar{\bar{\mathbf{V}}}$, where $\bar{\bar{\mathbf{V}}} \succeq 0$ and $\mathrm{rank}(\bar{\bar{\mathbf{V}}}) = 1$. The problem can then be transformed into an SDP by relaxing the rank constraint. Nonetheless, more favorable solutions can be obtained by further transforming the objective into an explicit optimization form \cite{Wu19}. To this end, we introduce the slack variables $\{\delta_{k}\}$ to denote the difference between the achievable values of the circuit constraints and their requirements:
\begin{subequations}
\begin{align}
\text{(M2.1)}: ~~\max_{\bar{\bar{\mathbf{V}}}, \{\delta_{k}\}} ~& \sum_{k=1}^{K} \delta_{k} 	\label{M21A} \\
\mathrm{s.t.}~~&\mathrm{tr}(\mathbf{U}_{k} \bar{\bar{\mathbf{V}}}) + |b_{k}|^{2} |d_{k}|^{2} \geq \frac{\gamma_{th} \sigma_{R}^{2}}{\alpha_{k} |b_{k}(t)|^{2}}, \ \forall k,  \label{M21B} \\
&\mathrm{tr}(\bar{\mathbf{C}}_{k} \bar{\bar{\mathbf{V}}}) + |d_{k}|^{2} \geq \frac{\xi + \delta_{k}}{1 - \alpha_{k}}, \ \forall k, 	 \label{M21C} \\
&\delta_{k} \geq 0, \ \forall k, \label{M21D} \\
&\bar{\bar{\mathbf{V}}}_{n, n} = 1, \ \forall n,  \label{M21E} \\
&\bar{\bar{\mathbf{V}}} \succeq 0.  \label{M21F}
\end{align}
\end{subequations}
Intuitively, this prioritizes the more limiting circuit constraint with the $\delta_{k}$ terms while ensuring the SNR constraints are also met. Problem (M2.1) can be readily solved using CVX. Again, Gaussian randomization is performed in each iteration to obtain a rank-one $\bar{\bar{\mathbf{V}}}$.\footnote{In choosing the randomization method, we compared the performance of all three methods (randA, randB, randC) in \cite{gaussrand}, and determined that while randA and randC performed near-identically, the quality of the candidate phase shift vectors both exceeded that from randB when the SNR constraint in (\ref{M21B}) is concerned.} Taking advantage of the slackness of the circuit constraint (\ref{M21C}), the randomization process selects the best candidate vector $\bar{\mathbf{v}}_{SDR}^{*}$ that results in the largest total SNR surplus for each tag over its requirement, which is translated into a reduction in $\left\lVert \mathbf{w} \right\rVert^{2}$ in the next iteration of Problem (M1.1). Each candidate vector $\bar{\mathbf{v}}_{SDR}$ is checked against constraints (\ref{M21B})-(\ref{M21C}), and the best $\bar{\mathbf{v}}_{SDR}^{*}$ is selected only from the candidates which meet both (\ref{M21B}) and (\ref{M21C}). As (\ref{M21A}) converges over the iterations of Problem (M2.1), the constraints in (\ref{M21B}) and (\ref{M21C}) are updated to narrow the solution space. \textbf{Algorithm 2} summarizes the procedure.

\begin{algorithm}
	\caption{MM-Based IRS Phase Shift Optimization with Multiple Tags}
	\begin{algorithmic}[1]
		\STATE \textbf{Initialize:} Beamforming vector $\mathbf{w}^{(i+1)}$, tag splitting coefficients $\bm{\alpha}^{(i)}$, combining vectors $\mathbf{G}^{(i)}$ where $i$ is the outer iteration number; a random starting point $\bar{\mathbf{v}}_{0}^{(j)}$; iteration number $j = 1$.
		\WHILE{the rate of change in objective function (\ref{M21A}) is above a threshold $\varepsilon > 0$}
			\STATE For each tag, construct $\mathbf{T}_{k}$ from $\bar{\mathbf{v}}_{0}^{(j)}$ and $\mathbf{U}_{k}$ from (\ref{U_k}).
			\STATE Solve Problem (M2.1) and denote the solution as $\bar{\bar{\mathbf{V}}}_{SDR}^{(j)}$.
			\STATE Perform eigenvalue decomposition $\bar{\bar{\mathbf{V}}}_{SDR}^{(j)} = \mathbf{\Lambda} \mathbf{D} \mathbf{\Lambda}^{H}$.
			\FOR{$q = 1$ to a required number of randomizations}
				\STATE Generate vector $\mathbf{r}_{q}$ according to the randA method in \cite{gaussrand}.
				\STATE Compute $\bar{\bar{\mathbf{v}}}_{q}\!\!=\!\!\bm{\Lambda} \mathbf{D}^{1/2} \mathbf{r}_{q}$, drop the last element to obtain $\bar{\mathbf{v}}_{SDR, q}$, and compute $F(\bar{\mathbf{v}}_{SDR, q})$.
			\ENDFOR
			\STATE Select the candidate vector resulting in the LHS of constraint (\ref{M21B}) exceeding the RHS by the largest amount over all $q$, and set $\bar{\mathbf{v}}_{0}^{(j)} \leftarrow \bar{\mathbf{v}}_{SDR}^{*(j+1)}$ for the next iteration.	
			\STATE Update inner iteration number $j \leftarrow j + 1$.
		\ENDWHILE
		\STATE \textbf{Return:} Optimized phase shift matrix $\mathbf{\Theta}^{*}$.
	\end{algorithmic}
\end{algorithm}

\textit{2) Successive Refinement Algorithm:} Each iteration of Problem (M2.1) incurs high complexity due to the SDP and randomization. Thus, we propose a novel low-complexity algorithm based on the SR approach in Section IV-B, with several modifications to cater for the multiple SNR and circuit constraints. The solution to Problem (M2.1) has the advantage that the residuals (i.e., the difference between the LHS and RHS) to both (\ref{M21B}) and (\ref{M21C}) are maximized: the latter through maximizing the slack variables $\{\delta_{k}\}$ in the objective, and the former through randomization, on the basis that the circuit constraints can already be met with high probability after maximizing $\sum_{k=1}^{K} \delta_{k}$. The proposed successive refinement scheme is designed to mimic the process of maximizing both SNR and circuit constraint residuals at once via a single objective. 

Similar to the single-tag case, in each SR iteration, the $N$ phase shifts are optimized sequentially starting from the first, while holding all others constant. Optimizing the $n$-th phase shift involves computing both the SNR and circuit constraint residuals for all $K$ tags over a vector of phase shift values $\theta_{n} = 0$:$T$:$2 \pi$, where $T$ is the phase shift precision. The resulting vectors are denoted by $S_{k}$ and $C_{k}$ for the $k$-th tag, respectively. For a certain $\theta_{n}$, the case where $S_{k}$ or $C_{k}$ is less than zero indicates that the corresponding constraint in Problem (M2.1) is violated. To construct the objective, we take the elementwise minima of the set of $\{S_{k}\}$ and $\{C_{k}\}$, to give us vectors of worst-case SNR and circuit constraint residuals over all tags. Then, both vectors are scaled such that their positive elements (if any) are normalized to the range $[0, 1]$. The post-scaling vectors are denoted by $S^{+}$ and $C^{+}$, respectively. Finally, the objective is defined as the elementwise product of $S^{+}$ and $C^{+}$. Intuitively, the value of $\theta_{n}$ that maximizes this objective helps to maximize the gap between the SNR and/or the available backscattering energy of the worst-performing tag and their requirements, while balancing over the better-performing tags. Where either $S^{+}$ or $C^{+}$ is all-zero, indicating that at least one tag does not meet the SNR or circuit constraints regardless of the phase shift used for the current reflector, the maximization over $\theta_{n}$ is performed with respect to the non-zero parts of the other vector. However, despite zero elements possibly being present in the objective, if Problem (M) was feasible before entering this step, this step will also provide a feasible solution.

The detailed procedure is described in \textbf{Algorithm 3}. To facilitate successive refinement with the circuit constraint, the LHS of the circuit constraint for tag $k$ can be written as a function of the $n$-th phase shift term $s_{n}$ in (\ref{Xisj}), in a similar manner as the LHS of the SNR constraint in (\ref{expand}).

\begin{figure*}
\normalsize
\begin{equation}
\Xi_{k}(s_{n}) = (1 - \alpha_{k}) \Bigg| [\mathbf{h}_{T_{k}I}^{H}]_{n} s_{n} [\mathbf{H}_{CI} \mathbf{w}]_{n} \\ + \sum_{j \neq n}^{N} [\mathbf{h}_{T_{k}I}^{H}]_{j} s_{j} [\mathbf{H}_{CI} \mathbf{w}]_{j} + \mathbf{h}_{CT_{k}} \mathbf{w} \Bigg|^{2}.	\label{Xisj}
\end{equation}
\hrulefill
\end{figure*}

\begin{algorithm}
	\caption{Successive Refinement-Based Multi-Tag IRS Phase Shift Optimization}
	\begin{algorithmic}[1]
		\STATE \textbf{Initialize:} Beamforming vector $\mathbf{w}^{(i+1)}$, phase shift matrix $\mathbf{\Theta}^{(i)}$, splitting coefficient vector $\bm{\alpha}^{(i)}$, combining vectors $\mathbf{G}^{(i)}$, vector of test phase shift values $\theta_{n} = 0$:$T$:$2 \pi$.
		\WHILE{the rate of change in the minimum SNR of all tags is above a threshold $\varepsilon > 0$}
			\FOR{$n$ = $1$:$N$}
				\FOR{$k$ = $1$:$K$}
					\STATE Compute $\alpha_{k} F(s_{n})$ in (\ref{expand2}) or (\ref{Fsj}) for all $\theta_{n}$ and subtract by $\gamma_{th} \sigma_{R}^{2}$ to obtain $S_{k}$.
					\STATE \textbf{if} $\xi > 0$: Compute $(1 - \alpha_{k}) \Xi(s_{n})$ in (\ref{Xisj}) for all $\theta_{n}$ and subtract by $\xi$ to obtain $C_{k}$.
				\ENDFOR
				\STATE Compute $\min \{S_{k}\}$ and scale its positive portion to $[0, 1]$ to obtain $S^{+}$.
				\STATE \textbf{if} $\xi > 0$: Compute $\min \{C_{k}\}$ and scale its positive portion to $[0, 1]$ to obtain $C^{+}$.
				\STATE \textbf{if} $\xi = 0$: Set $\theta_{n}^{*} = \mathrm{arg} \max_{\theta} S^{+}$;
				\STATE \textbf{else if} $S^{+} \odot C^{+}$ has nonzero elements: Set $\theta_{n}^{*} = \mathrm{arg} \max_{\theta} S^{+} \odot C^{+}$;
				\STATE \textbf{else if} $S^{+} \odot C^{+} = \mathbf{0}^{T}$ but $S^{+}$ is nonzero: Set $\theta_{n}^{*} = \mathrm{arg} \max_{\theta} C^{+}$;
				\STATE \textbf{else:} Set $\theta_{n}^{*} = \mathrm{arg} \max_{\theta} S^{+}$.
			\ENDFOR
		\ENDWHILE
		\STATE \textbf{Return:} Optimized phase shift matrix $\mathbf{\Theta}^{*}$.
	\end{algorithmic}
\end{algorithm}

\subsection{Tag Splitting Coefficient and Receive Beamforming Optimization}

Next, the feasibility problem for the splitting coefficients of all tags can be written as
\begin{subequations}
\begin{align}
\text{(M3)}: ~~\mathrm{find} ~~&\bm{\alpha} 		\label{M3A} \\
\mathrm{s.t.}~~&\left| \mathbf{g}_{k}^{H} \mathbf{h}_{k, 2}(\mathbf{\Theta}) \right|^{2} \mathrm{tr}(\mathbf{W} \mathbf{H}_{k, 1}(\mathbf{\Theta})) \nonumber \\
& \qquad \qquad \qquad \quad \geq \frac{\gamma_{th} \sigma_{R}^{2}}{\alpha_{k} |b_{k}(t)|^{2}}, \ \forall k, 	  \label{M3B} \\
&\mathrm{tr}(\mathbf{W} \mathbf{H}_{k, 1}(\mathbf{\Theta})) \geq \frac{\xi}{1 - \alpha_{k}}, \ \forall k, 	\label{M3C} \\
&0 \leq \alpha_{k} \leq 1, \ \forall k. 	\label{M3D}
\end{align}
\end{subequations}
Noting that (\ref{M3B}) and (\ref{M3C}) can be combined with $\alpha_{k}$ on the LHS, (\ref{M3D}) is equivalent to
\begin{multline}
\frac{\gamma_{th} \sigma_{R}^{2}}{|b_{k}(t)|^{2} \mathrm{tr}(\mathbf{W} \mathbf{H}_{k, 1}(\mathbf{\Theta})) \left| \mathbf{g}_{k}^{H} \mathbf{h}_{k, 2}(\mathbf{\Theta}) \right|^{2}} \leq \alpha_{k} \\ \leq 1 - \frac{\xi}{\mathrm{tr}(\mathbf{W} \mathbf{H}_{k, 1}(\mathbf{\Theta}))}.	\label{modifiedQ4D}
\end{multline}
Constraint (\ref{modifiedQ4D}) denotes the range of feasible splitting coefficients, where any value in this range suffices. As the residual in either the SNR or the circuit constraint is reduced over the $\mathbf{w}$ and $\mathbf{\Theta}$ sub-problems, the range of feasible $\alpha_{k}$ converges to a single feasible point over the iterations.

Finally, given orthogonal tag transmissions, the combiner which maximizes the strength of the received signal over all antennas for each tag is given by $\mathbf{g}_{k}^{*} = \frac{\mathbf{h}_{k, 2}(\mathbf{\Theta}) \mathbf{h}_{k, 1}(\mathbf{\Theta}) \mathbf{w}}{\left\lVert \mathbf{h}_{k, 2}(\mathbf{\Theta}) \mathbf{h}_{k, 1}(\mathbf{\Theta}) \mathbf{w} \right\rVert}$.

\subsection{Overall Algorithm and Complexity Analysis}

The procedure to solve Problem (M) is summarized in \textbf{Algorithm 4}.

\begin{algorithm}
	\caption{Alternating Optimization (AO) Algorithm for Multi-Tag Systems}
	\begin{algorithmic}[1]
		\STATE \textbf{Initialize:} Random IRS phase shifts $\mathbf{\Theta}^{(1)}$; random splitting coefficients $\bm{\alpha}^{(1)}$; random combining vectors $\mathbf{G}^{(1)}$; set iteration number $i = 1$.
		\WHILE{the rate of change of objective function (\ref{M1A}) is above a threshold $\varepsilon > 0$} 
			\STATE Solve Problem (M1.1) using $\mathbf{\Theta}^{(i)}$, $\bm{\alpha}^{(i)}$ and $\mathbf{G}^{(i)}$ and denote the solution as $\mathbf{W}_{SDR}^{(i)}$.
			\STATE Perform Gaussian randomization as per \cite{gaussrand} to obtain the best solution $\mathbf{w}^{(i+1)}$.
			\STATE Solve Problem (M2.1) using either \textbf{Algorithm 2} or \textbf{Algorithm 3} to obtain $\mathbf{\Theta}^{(i+1)}$.
			\STATE Solve Problem (M3) using $\mathbf{W}^{(i+1)}$, $\bm{\Theta}^{(i+1)}$ and $\mathbf{G}^{(i)}$ by selecting $\bm{\alpha}^{(i+1)}$ to be within the bounds in (\ref{modifiedQ4D}).
			\STATE Obtain $\mathbf{G}^{(i+1)}$ by computing the optimal combiner $\mathbf{g}_{k}^{*}$ for all $k$.
			\STATE Update iteration number $i \leftarrow i + 1$.
		\ENDWHILE
		\STATE \textbf{Return:} Minimized transmit power, $P^{*} = \left\lVert \mathbf{w}^{*} \right\rVert^{2}$.
	\end{algorithmic}
\end{algorithm}

While the relaxed SDPs in Problems (M1.1) and (M2.1) mean that optimality to \textbf{Algorithm 4} cannot be guaranteed, the objective value is nonincreasing in each iteration. This is due to the fact that the solutions from subproblems (M2.1) and (M3) are then converted to a reduction in the transmit power in the subsequent iteration of Problem (M1.1). Thus, where feasible solutions are found in both Problems (M2.1) and (M3), $\left\lVert \mathbf{w} \right\rVert^{2}$ decreases in the subsequent iteration.

\begin{table}
\caption{Computational complexities of the components of the multi-tag algorithms.}
\label{complexities}
\setlength{\tabcolsep}{3pt}
\centering
\begin{tabular}{|c|c|}
\hline
\textbf{Step} & Complexity \\
\hline
Tx. beamforming ($\mathbf{w}$) & \begin{tabular}{@{}c@{}} $O(\max\{K, L\}^{4} L^{1/2})$ for SDP \cite{Luo10a}, \\ $O(RN^{2})$ for randomization \end{tabular} \\
\hline
Phase shifts ($\mathbf{\Theta}$) & \begin{tabular}{@{}c@{}}MM-based (\textbf{Algorithm 2}): \\ $O(I_{MM} (\max\{2K,\!N\}^{4} N^{1/2}\!+\!N^{3}\!+\!KRN^{2}))$ \\ SR-based (\textbf{Algorithm 3}): \\ $O(I_{SR} KN^{2}/T)$ \end{tabular} \\
\hline
Splitting coeff. ($\bm{\alpha}$) & $O(KL^{3} + KN^{2})$ \\
\hline
Rx. beamforming ($\mathbf{G}$) & $O(K N^{2})$ \\
\hline
\end{tabular}
\end{table}

The complexities of the transmit beamforming optimization, phase shift optimization, tag splitting coefficient computation and the receive beamforming optimization per AO iteration are shown in Table~I; the typical convergence behavior is discussed in Section VI-C. For the general case, $I_{MM}$ and $I_{SR}$ refer to the number of MM sub-iterations in Problem (M2.1) and the number of SR cycles through all IRS phase shifts in lines 3-14 of \textbf{Algorithm 3}, respectively; $R$ is the number of randomization iterations. For the MM-based phase shift optimization, the three terms are due to the SDP in Problem (M2.1) (where $\bar{\bar{\mathbf{V}}}$ is a square matrix of size $N+2 \approx N$), the eigenvalue decomposition after solving the SDP, and randomization, respectively. For the SR-based phase shift optimization, $T$ is the precision of the linear search over each phase shift, where $\left\lceil \frac{2 \pi}{T} \right\rceil$ is the number of discrete phase shift values in the vector $\theta_{n}$ in \textbf{Algorithm 3}.  These are asymptotic complexities based on the assumptions that $L, M \ll N$ and $R \gg N$. Moreover, when tags are semi-passive, Problems (M1.1) and (M2.1) are solved in the absence of constraints (\ref{M11C}) and (\ref{M21C}), respectively, and Problem (M3) and its associated computational cost is omitted.


\section{Numerical Results}

In this section, we numerically quantify the transmit power reduction at the CE in the IRS-aided bistatic BackCom system. We begin with the performance of the base case algorithms in Section IV in a system with a single semi-passive tag, and extend the results to a single passive tag in a special case of the general algorithm in Section V. Then, we quantify the performance of the general algorithm in a multi-tag system where tags may be either semi-passive or passive, and examine the convergence characteristics of our proposed algorithms.

\begin{figure*}[h!]
	\centering
	\includegraphics[width=0.95\textwidth]{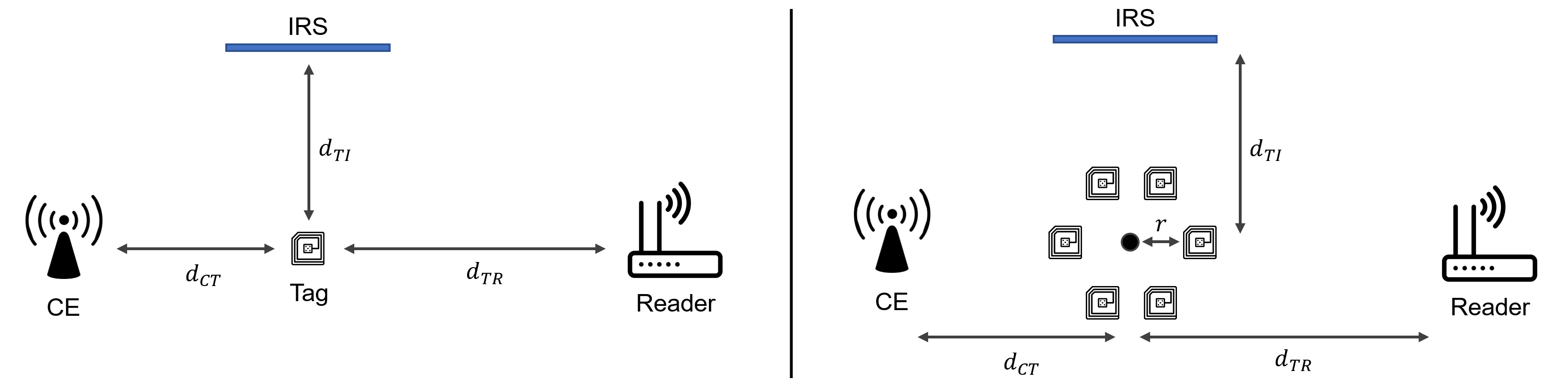}
	\caption{Simulation setup for (left) the single-tag setup and (right) the multi-tag setup.}
	\label{fig:simulationSetup}
\end{figure*}

The CE is located at the origin and the reader is located at $[100, 0]$ unless otherwise noted, with all coordinates in meters hereafter. For the single-tag system, the CE has $L = 4$ antennas, and the tag is located on a straight line between the CE and the reader. For the multi-tag system, $L = 4$ and the reader also has $M = 4$ antennas, and $K$ tags are centered around $[20, 0]$ at a distance of $5$ m, with equal angular spacing to simulate a cluster of tags. The default location of the IRS is $[20, 20]$. Fig.~\ref{fig:simulationSetup} shows the system setup for both the single- and multi-tag systems.

The CE transmits at the frequency of $915$ MHz, typical of the RFID band \cite{GD09}. We assume a square IRS and consider an outdoor scenario, where all links undergo Rician fading with $K$-factor $3$ dB and path loss exponent $2.1$. The Rician $K$-factor is drawn from the typical range in \cite{Alev18} after conversion from the Nakagami parameter; while the path loss exponent is chosen to be higher than the free-space assumption in \cite{Kim14} to account for sparse scatterers in the environment. The distances between the CE, tag and the reader are typical of bistatic BackCom systems \cite{Kim14}, with the tag being slightly further from the reader in our case to demonstrate the improved tag deployment flexibility as a result of the IRS. Unless otherwise noted, the number of channel realizations in all simulations is $1000$; the default number of IRS elements is $N = 64$; the tag's baseband signal magnitude is $|b(t)|^{2} = 1$; the SNR requirement for all tags is $\gamma_{th} = 8$ dB; the noise power at the reader is $\sigma_{R}^{2} = -110$ dBm; the Lipschitz constant is set to $\ell = 2.5 \times 10^{-16}$ and the convergence threshold for the AO-based algorithms is set to $\varepsilon = 10^{-4}$. The default IRS size is a representative value in the range used in \cite{Wu19}. The tag's SNR requirement of $8$ dB is based on \cite{Kim14}, which resulted in a BER of $0.01$ in the bistatic system therein. The Lipschitz constant is chosen via extensive simulations to cater to the objective function $F(\bar{\mathbf{v}})$ based on a range of tag locations and associated channel gains, and is used throughout this section; while the convergence threshold is a stricter version of the $10^{-3}$ used in \cite{Wu19}.

We adopt the path loss model in \cite[Eq. (23)]{Ell19}, which is applicable to both near- and far-field transmissions involving an IRS reflection. The path losses for the $C$-$I$, $T$-$I$ and $R$-$I$ links are split into their respective components based on \cite[Eq. (21)-(22)]{Ell19}, and are directly absorbed into $\mathbf{h}_{TI}^{H}$, $\mathbf{H}_{RI}^{H}$ and $\mathbf{H}_{CI}$. The path losses for the $C$-$T$ and $T$-$R$ links are directly absorbed into $\mathbf{h}_{CT}$ and $\mathbf{h}_{CR}$, respectively, consistent with \cite[Eq. (22)]{Ozd19}. 

\subsection{Single Semi-Passive Tag with Single-Antenna Reader}

In this subsection, the minimized CE transmit powers are obtained by solving Problem (S) using both \textbf{Algorithm 1} (i.e., the MM algorithm) in Section IV-A and the successive refinement scheme in Section IV-B (i.e., the SR algorithm).

\begin{figure}[h!]
	\centering
	\includegraphics[width=3.5in]{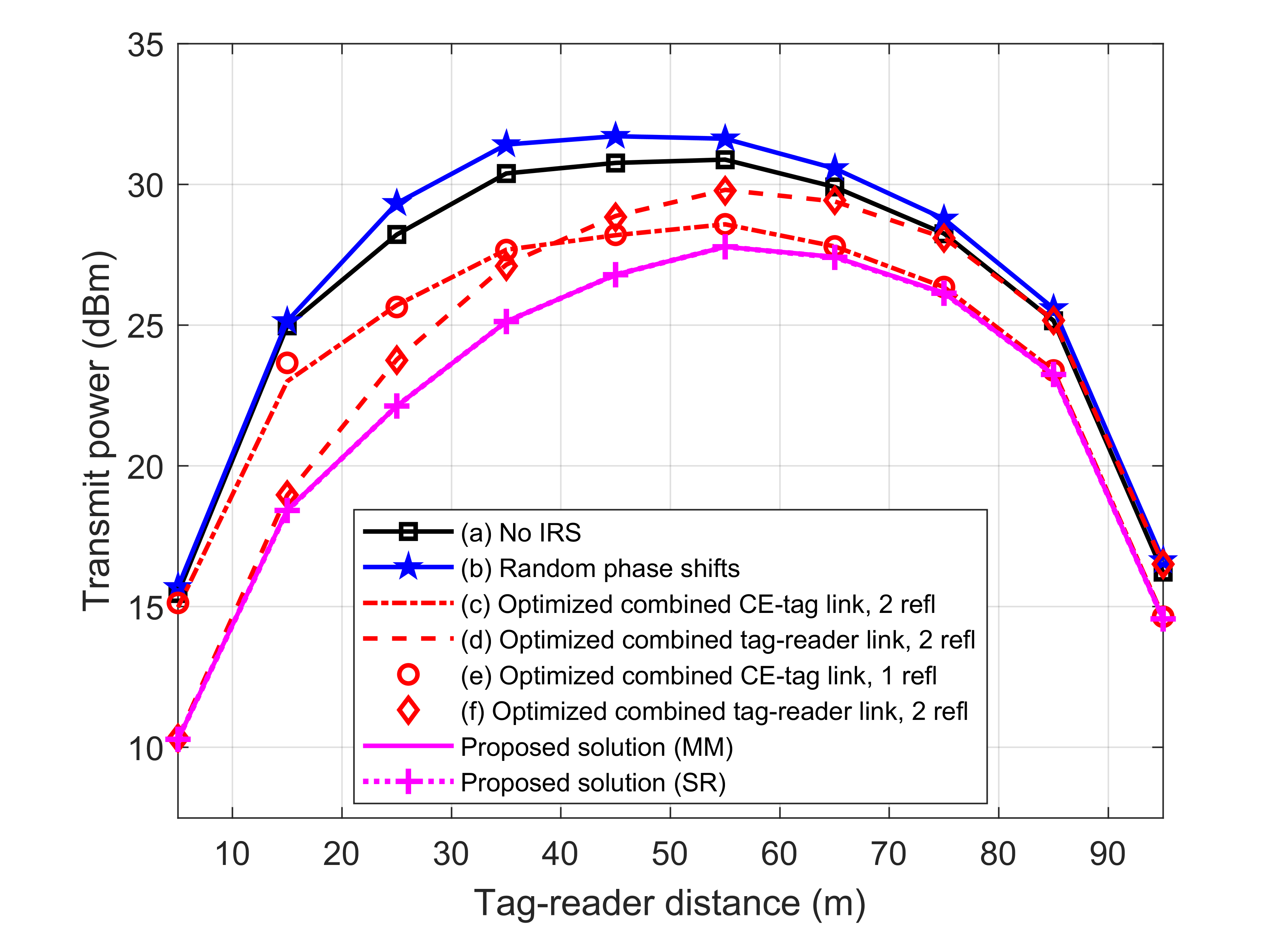}
	\caption{Effect of tag location on the transmit power at the CE.}
	\label{fig:singleUserTagLocation}
\end{figure}

\textit{1) Effect of Tag Location:} Fig. \ref{fig:singleUserTagLocation} shows the minimum CE transmit power as the tag is moved along a straight line between $[5, 0]$ and $[95, 0]$. In addition to our proposed algorithms, several suboptimal baseline schemes are adopted as benchmarks. These include: (a) no-IRS (where MRT is utilized for transmit beamforming); (b) random IRS phase shifts; (c) using a set of suboptimal phase shifts for the combined CE-tag link only (i.e., $C$-$T$ plus $C$-$I$-$T$); (d) using the optimal phase shifts for the combined tag-reader link only (i.e., $T$-$R$ plus $T$-$I$-$R$)\footnote{The phase shifts used in benchmark (c) (i.e., combined CE-tag link) are suboptimal, as they are obtained in an alternating manner over $\mathbf{w}$ and $\mathbf{\Theta}$ similar to the single-user case in \cite{Wu19}. On the other hand, the phase shifts used in benchmark (d) (i.e., combined tag-reader link) are optimal, and are given by $\theta_{n}^{*} = \theta_{TR} - \theta_{RI_{n}} - \theta_{TI_{n}}$, where $\theta_{TR}$, $\theta_{RI_{n}}$ and $\theta_{TI_{n}}$ are the phases of the tag-reader, individual IRS element-reader and IRS element-tag channels, respectively. The method used to obtain benchmark (c) was shown in \cite{Wu19} to perform identically to an SDR-based algorithm, and hence suffices as a near-optimal benchmark.}, and (e), (f) as single-reflection variants of (c) and (d). It is clear that notable power reductions are realized at all locations compared to the no-IRS system --- up to $6$ dB when the tag is closer to the CE. Moreover, the optimized transmit power is around $27$ dBm or less for all tag locations, which enables much-improved tag placement flexibility using reasonable transmit powers. The MM and SR algorithms perform similarly, confirming their convergence behavior. We also note that the two suboptimal schemes (c) and (d) that phase-align to either the combined CE-tag link or the combined tag-reader link is close to that of \textbf{Algorithm~1} when the tag is near the reader or the CE, respectively. The extent of power reduction is not symmetric with respect to tag location, suggesting that the IRS location is also of influence. Contrary to the observations of works on conventional systems (e.g., \cite{Wu19}), implementing random phase shifts at the IRS appears to hinder the balance between the two links, due to the presence of deterministic components in the channels. As the random phase scheme does not outperform the no-IRS benchmark over the range of tag locations in Fig.~\ref{fig:singleUserTagLocation}, it is unlikely to be comparable to any of the algorithms proposed in this paper, and is thus excluded from subsequent results.

\textit{2) Importance of Twice-Reflection in the Signal Model:} Noting that \cite{Chen21} considered only one reflection at the IRS for a signal from the transmitter to receiver, we also evaluated the effect of the single-reflection signal model on the transmit power, through benchmarks (e) and (f). Therein, the IRS is optimized with respect to the combined CE-tag link only and the combined tag-reader only, respectively, where the other combined link is not included in the signal model. One may observe that benchmarks (e) and (f) perform almost identically compared to benchmarks (c) and (d), respectively, but still substantially worse than the solutions from the MM/SR algorithms. This highlights that both the combined CE-tag and tag-reader links contribute significantly to the overall performance of the system when the IRS is optimized with respect to both combined links; thus, it is important to include the twice-reflection path in the signal model. Moreover, the consideration of only one reflection at the IRS, or optimizing the IRS with respect to only one of the two combined links, leads to considerable performance degradation.

\begin{figure}[h!]
	\centering
	\includegraphics[width=3.5in]{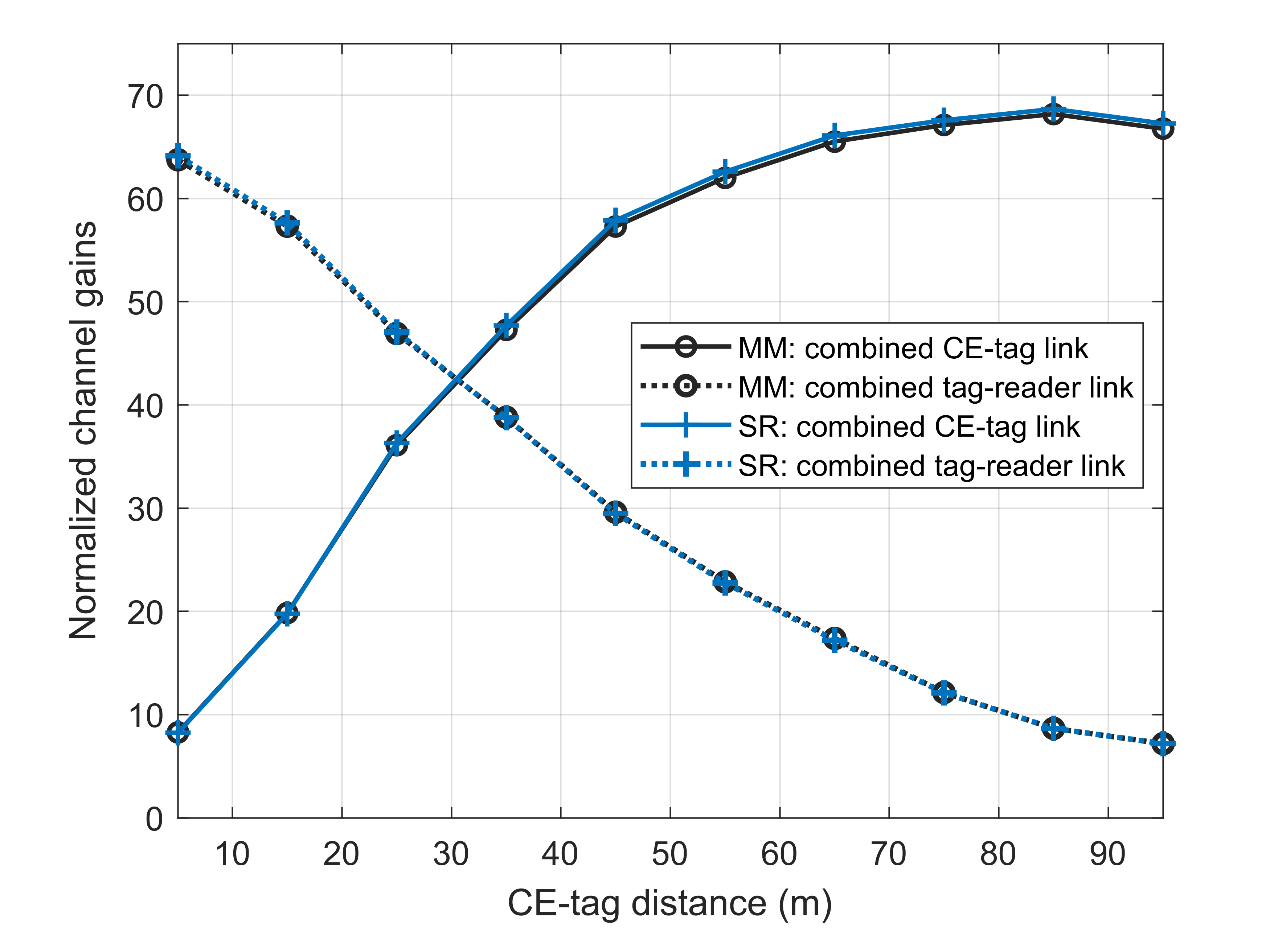}
	\caption{Normalized gains of the combined CE-tag and tag-reader links.}
	\label{fig:singleUserChannelSums}
\end{figure}

\textit{3) Behavior of IRS Phase Shifts:} To gain insight into the phase shift configuration of the IRS when a signal is reflected twice, Fig. \ref{fig:singleUserChannelSums} plots the normalized channel gains of the combined CE-tag link and the combined tag-reader link, where the IRS phase shifts are tuned according to the solution from the MM and SR algorithms. The normalized channel gain is a path loss-independent measure of the extent to which the phase shifts of the IRS are aligned with a given channel. They are given by $\hat{h}_{CIT} = \left| \left( \hat{\mathbf{h}}_{TI}^{H} \mathbf{\Theta} \hat{\mathbf{H}}_{CI} + \hat{\mathbf{h}}_{CT} \right) \mathbf{w} \right|$ and $\hat{h}_{TIR} = \left| \hat{\mathbf{h}}_{RI}^{H} \mathbf{\Theta} \hat{\mathbf{h}}_{TI} + \hat{h}_{TR} \right|$, respectively, for the combined CE-tag and tag-reader links, where the $\hat{\cdot}$ quantities denote their respective channel matrices with all elements set to unit magnitude. From Fig.~\ref{fig:singleUserChannelSums}, we observe that the phase shifts obtained with our algorithms generally boost the weaker link. For example, when the tag is closer to the CE, the IRS phase shifts are configured to favor the combined tag-reader link more than the combined CE-tag link. The intersection of the curves occurs when neither combined link is more favorable than the other, in this case when the tag is located $30$ m from the CE. At this point, the IRS achieves a balance between the two combined links.

\begin{figure}[h!]
	\centering
	\begin{subfigure}{0.495\textwidth}
		\centering
		\includegraphics[width=3.5in]{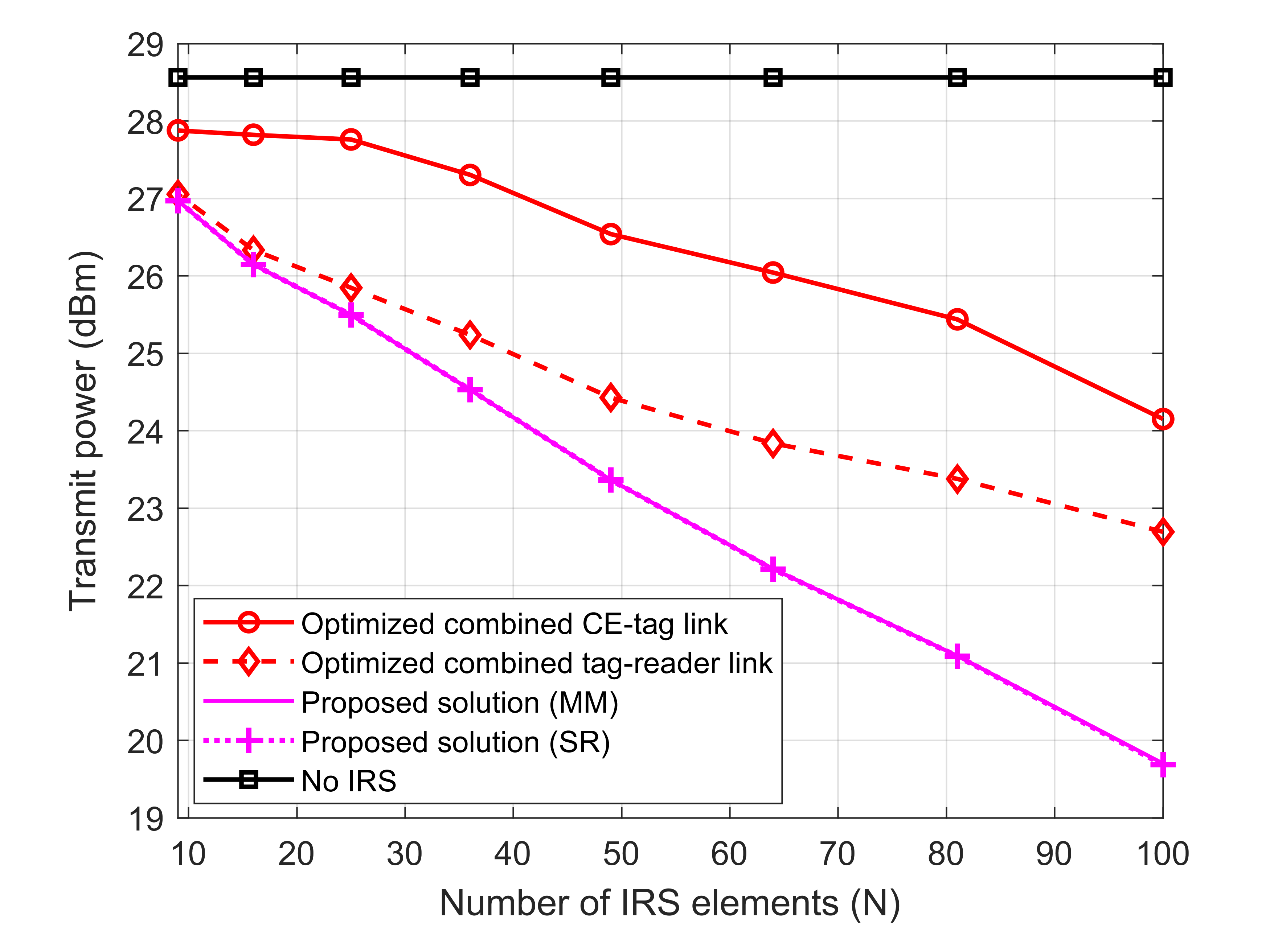}
		\caption{Minimum transmit power vs. $N$}
	\end{subfigure} \hfill
	\begin{subfigure}{0.495\textwidth}
		\centering
		\includegraphics[width=3.5in]{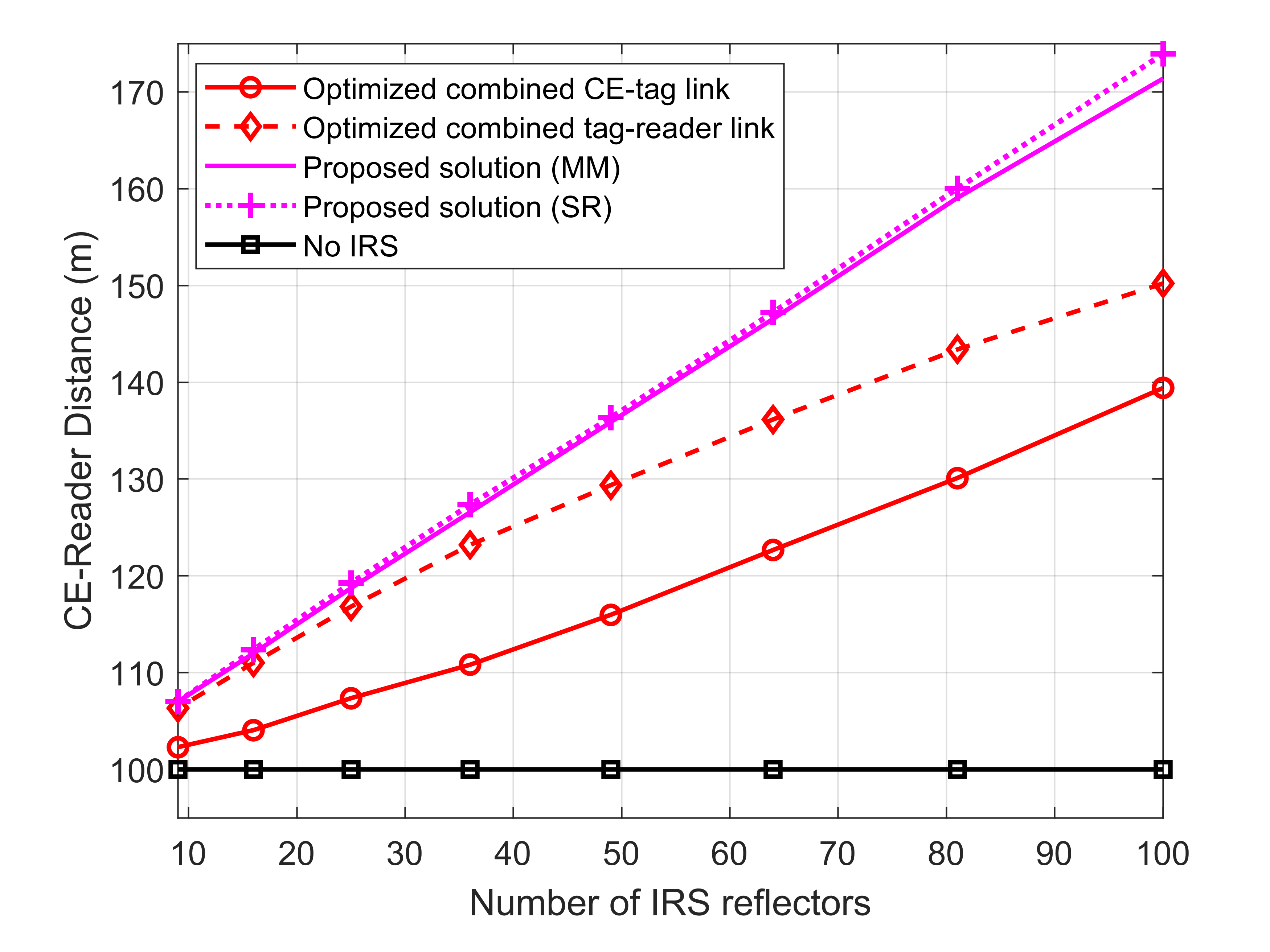}
		\caption{Range improvement using baseline transmit power}
	\end{subfigure}
	\caption{Effect of the number of IRS elements on transmit power and range.}
	\label{fig:singleUserN}
\end{figure}

\textit{4) Effect of the Number of IRS Elements:} Fig. \ref{fig:singleUserN}(a) compares the transmit power at the CE against the number of IRS elements, with the tag located at $[25, 0]$. The reduction in transmit power is roughly proportional to the number of IRS elements for both MM and SR algorithms and the suboptimal benchmarks compared to the no-IRS system. For a medium-sized IRS with $N = 64$, an average reduction of $6$ dB is achieved, increasing to over $8$ dB when $N = 100$.

The addition of the IRS allows considerable improvements in the link budget. Suppose the tag-reader distance is increased by moving the reader further. As the path loss of the combined CE-tag link is unchanged, the overall path loss from CE to reader is given in (\ref{pathlossEqnNo})-(\ref{pathlossEqn}), 
\begin{figure*}[!t]
\normalsize
\begin{subequations}
\begin{align}
\text{No-IRS: } \mathrm{PL} &= \left( \frac{\lambda}{4 \pi} \right)^{4} \frac{1}{d_{CT}^{\delta} (d_{TR} + \Delta)^{\delta}} , \label{pathlossEqnNo} \\
\text{IRS-aided: } \mathrm{PL} &= \left( \frac{\lambda}{4 \pi} \right)^{4} \Biggl[\frac{1}{d_{CT}^{\delta}} + \left( \frac{\lambda}{4 \pi} \right)^{2} \left| \sum_{n=1}^{N} e^{j \theta_{n}} \sqrt{\frac{\pi^{2} (-\hat{\mathbf{r}}_{CI,n} \cdot \hat{\mathbf{n}})^{2q} (\hat{\mathbf{r}}_{TI,n} \cdot \hat{\mathbf{n}})^{2q}}{d_{CI, n}^{\delta} d_{TI, n}^{\delta}}} \right|^{2} \Biggr] \nonumber \\
		& \qquad \times \Biggl[\frac{1}{(d_{TR} + \Delta)^{\delta}} + \left( \frac{\lambda}{4 \pi} \right)^{2} \left| \sum_{n=1}^{N} e^{j \theta_{n}} \sqrt{\frac{\pi^{2} (-\hat{\mathbf{r}}_{TI,n} \cdot \hat{\mathbf{n}})^{2q} (\hat{\mathbf{r}}_{RI,n} \cdot \hat{\mathbf{n}})^{2q}}{d_{TI, n}^{\delta} (d_{RI, n} + \Delta_{n})^{\delta}}} \right|^{2} \Biggr], \label{pathlossEqn}
\end{align}
\end{subequations}
\hrulefill
\end{figure*}
where $\Delta$ and $\Delta_{n}$ are the changes in the distances between tag and reader, and the $n$-th IRS element and the reader, respectively; $q$ is a constant defined in \cite{Ell19}; $\hat{\mathbf{r}}_{\cdot, n}$ represents the unit vector between two nodes; and $\hat{\mathbf{n}}$ is the vector normal to the IRS surface. While it is difficult to solve for the range increase $\Delta$ relative to the no-IRS system directly from the above equations, it can be easily computed numerically. The achievable ranges are visualized in Fig. \ref{fig:singleUserN}(b), where the tag is located at $[25, 0]$ and the transmit power is held constant for both the IRS-aided and non-IRS-aided systems (where $\mathbf{w}$ is MRT) at the minimized transmit power in a non-IRS-aided system. The approximately linear relationship between $N$ and the CE-reader distance is due to the square scaling behavior of the SNR in an IRS-assisted system and inverse-square scaling of the large-scale path loss. In any case, the range increases enabled by the IRS are significant, ranging from $12$ m for a small IRS with $N = 16$, to around $70$ m when $N = 100$.

\subsection{Single Passive Tag with Circuit Power Constraint}

\begin{figure}[h!]
	\centering
	\begin{subfigure}{0.495\textwidth}
		\centering
		\includegraphics[width=3.5in]{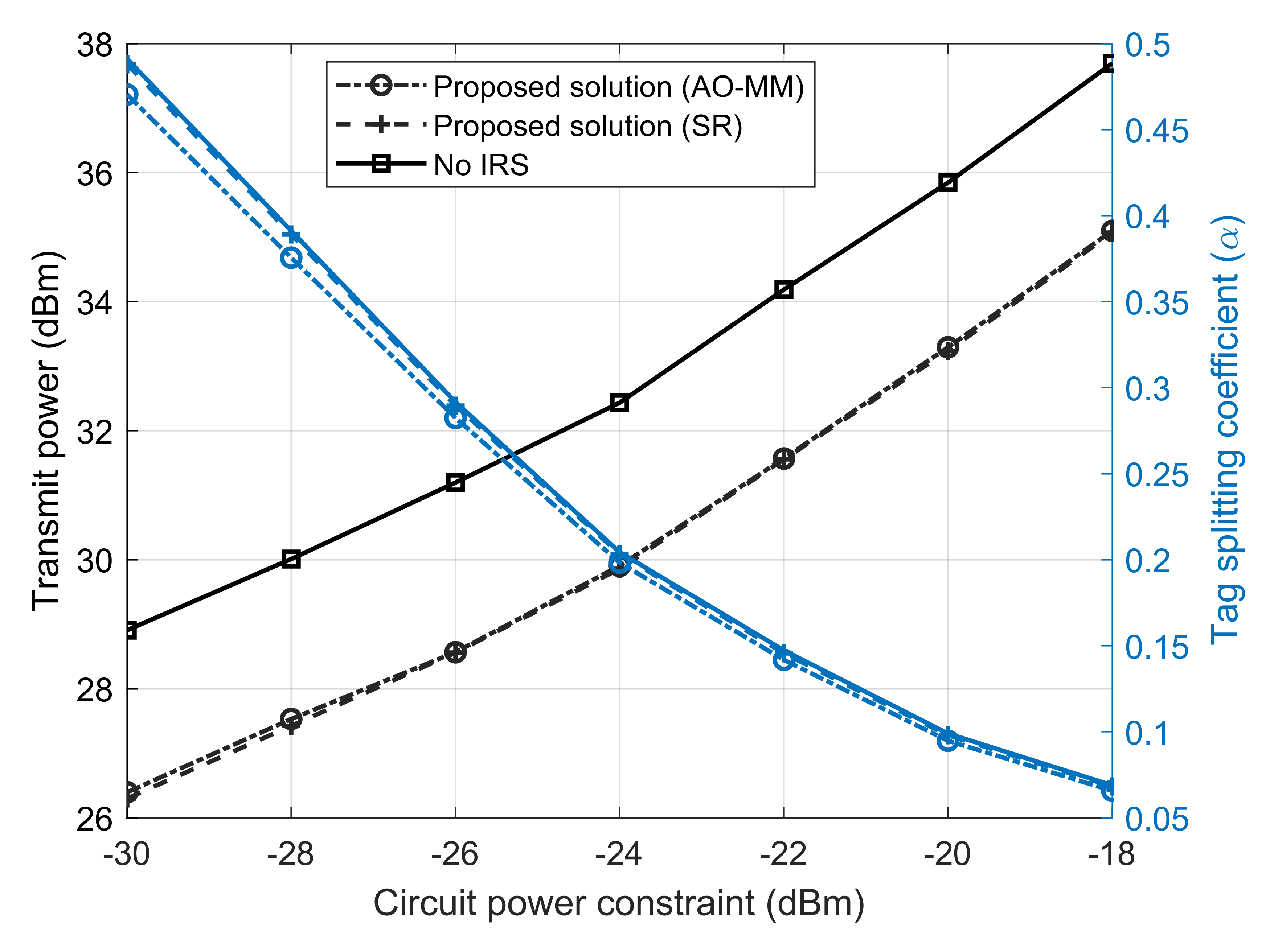}
		\caption{Varying $\xi$}
	\end{subfigure} \hfill
	\begin{subfigure}{0.495\textwidth}
		\centering
		\includegraphics[width=3.5in]{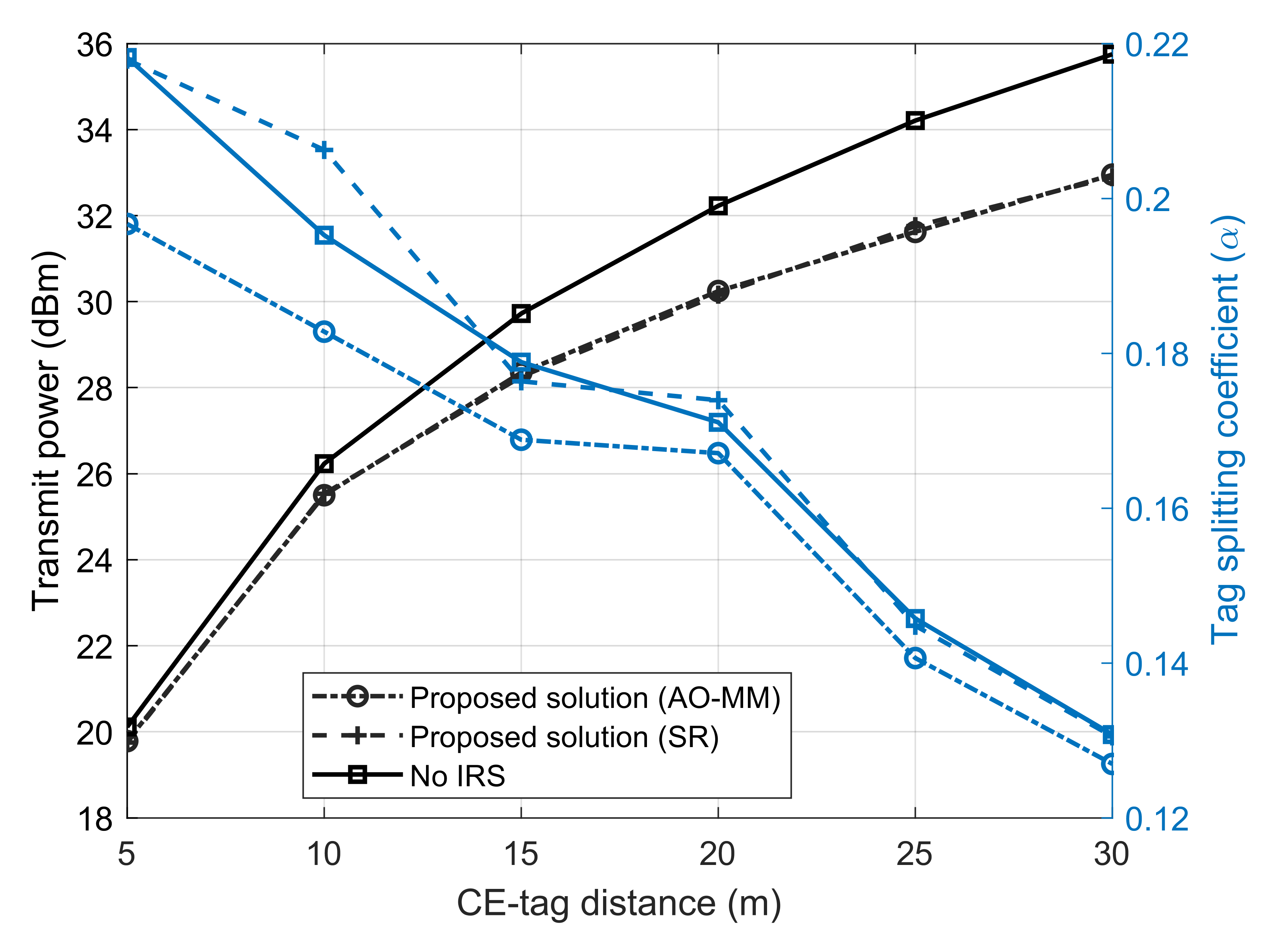}
		\caption{Varying CE-tag distance}
	\end{subfigure}
	\caption{Effect of circuit power constraint on the transmit power at the CE.}
	\label{fig:singleUserCPS}
\end{figure}

Next, we extend the numerical analysis to a passive tag, and examine the effects of a nonzero circuit power constraint on the CE's transmit power. As a passive tag falls under the scope of the general problem, the CE's transmit power is obtained using both the MM and SR variants of the AO algorithm (\textbf{Algorithm 4}) with the number of tags set to $1$. Fig.~\ref{fig:singleUserCPS}(a) highlights the effects of various tag circuit power constraint values on the transmit power. Compared to the results from Fig. \ref{fig:singleUserTagLocation}, a nonzero tag power requirement significantly increases the transmit power, as the tag cannot communicate, let alone achieve its SNR requirement, if it does not power on. The extent of power reduction relative to the no-IRS system is less than the semi-passive tag scenario, at around $2.5$ dB. However, in absolute terms, several Watts are conserved at the maximum $\xi$, which is of significant benefit if the CE is subject to transmit power regulations. Also, for practical values of $\xi$ shown in Fig. \ref{fig:singleUserCPS}(a), an order-of-magnitude increase in $\xi$ results in $7$-$8$ dB increase in the transmit power, indicating a nonlinear relationship. The optimal tag splitting coefficient decreases with increased $\xi$, such that a smaller proportion of the incoming signal is backscattered bearing the tag's data symbols. This requires the CE transmit power to be increased to meet both the SNR and circuit constraints.

Fig. \ref{fig:singleUserCPS}(b) highlights the effect of the circuit power constraint, with $\xi = -22$ dBm, as the tag location is varied between $[5, 0]$ and $[30, 0]$, which is further than the typical tag-reader distances in bistatic BackCom systems even with semi-passive tags. Compared to Fig. \ref{fig:singleUserTagLocation}, although the power reduction is diminished at all distances, the effect is still non-trivial: around $3$ dB reduction when the CE-tag distance is $30$ m, corresponding to more than $3$ W in real terms. The variations in the tag splitting coefficients can be attributed to the small-scale fading in the CE-tag link and the alternating optimization algorithm in dealing with variations in channel coefficients, but a diminishing trend is nonetheless observed as the tag is further removed from the CE.

\subsection{Multiple Tags with Multiantenna Reader}

In this subsection, we present the results from the general problem based on \textbf{Algorithm 4} in a system with multiple tags, where the tags are either semi-passive or passive.

\begin{figure}[h!]
	\centering
	\begin{subfigure}{0.495\textwidth}
		\centering
		\includegraphics[width=3.5in]{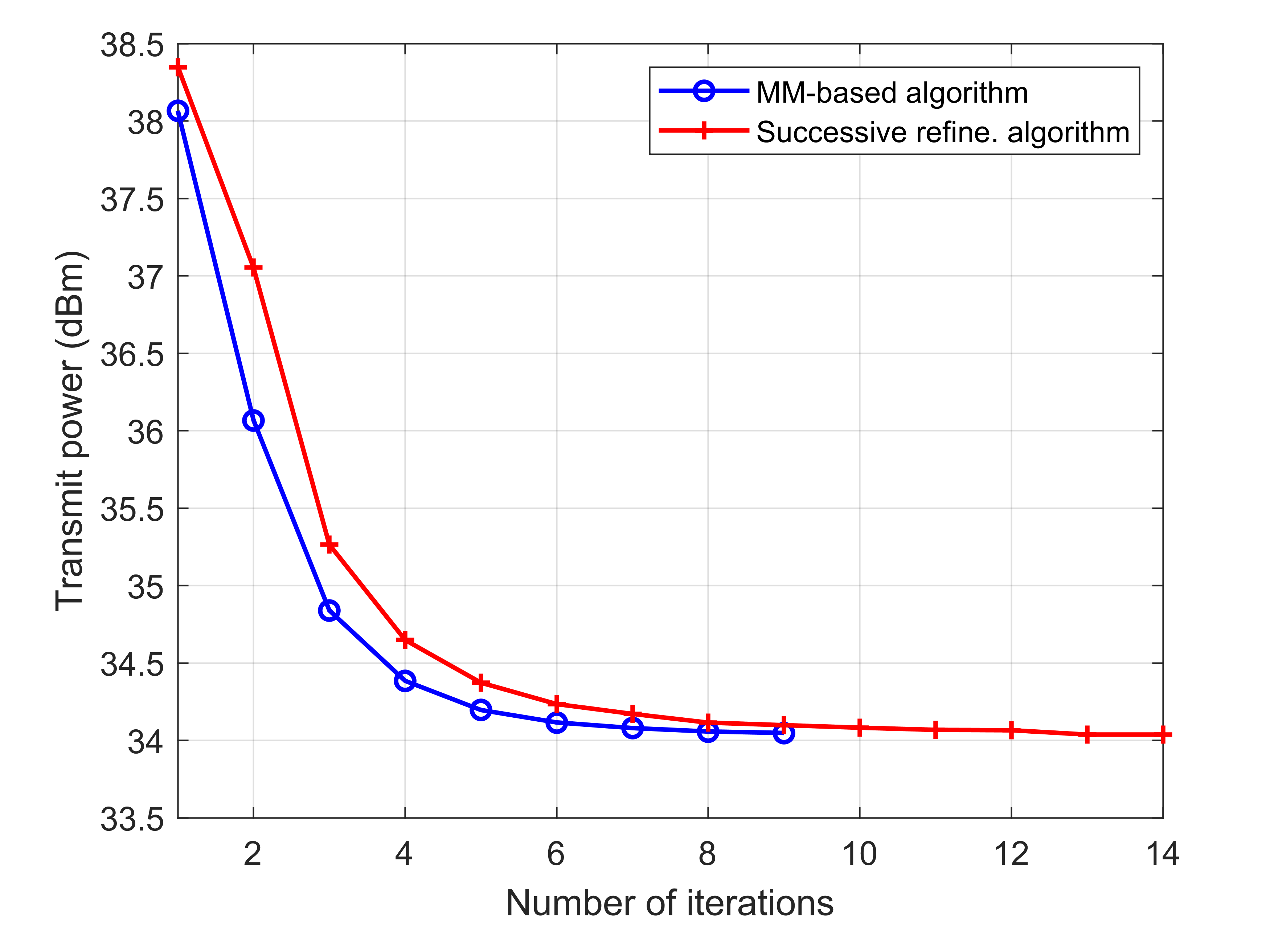}
		\caption{Convergence behavior of Algorithms 2 and 3}
	\end{subfigure} \hfill
	\begin{subfigure}{0.495\textwidth}
		\centering
		\includegraphics[width=3.5in]{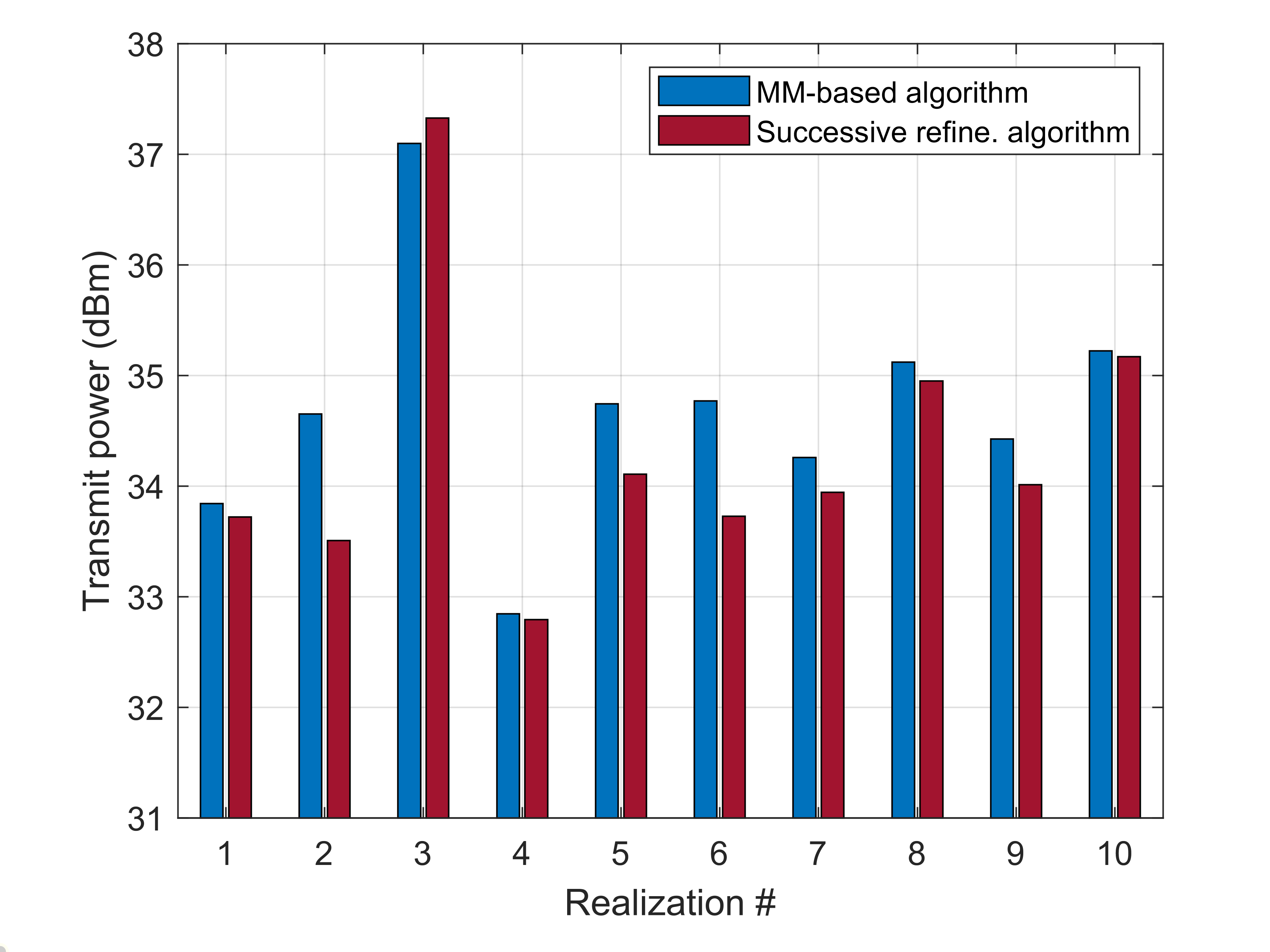}
		\caption{Comparison of converged objective values}
	\end{subfigure}
	\caption{Convergence comparison of the MM-based and successive refinement-based algorithms.}
	\label{fig:multiUserConvergence}
\end{figure}

\textit{1) Convergence Comparison of Algorithms:} Fig. \ref{fig:multiUserConvergence}(a) shows the typical convergence behavior of the multi-tag transmit power minimization algorithm using AO and either \textbf{Algorithm 2} or \textbf{Algorithm 3} for the IRS phase shift optimization, with $K = 6$. Both the MM-based algorithm and the SR scheme are capable of converging to similar solutions. While the latter algorithm generally requires more iterations to converge, it is capable of doing so with far lower computational cost than the MM-based algorithm, as discussed in Section V-D.

Due to varying channel coefficients, the converged value of $\left\lVert \mathbf{w} \right\rVert^{2}$ exhibits considerable variation between channel realizations. Fig. \ref{fig:multiUserConvergence}(b) plots the converged objective values of \textbf{Algorithm 4} with both the MM-based approach using \textbf{Algorithm 2} and the SR scheme using \textbf{Algorithm 3}, over $10$ representative channel realizations, where $\xi = -22$ dBm. While the SR scheme does not outperform the MM-based algorithm all the time, the typical difference in the converged objective values is reasonably small in most cases.

\begin{figure}[h!]
	\centering
	\begin{subfigure}{0.495\textwidth}
		\centering
		\includegraphics[width=3.5in]{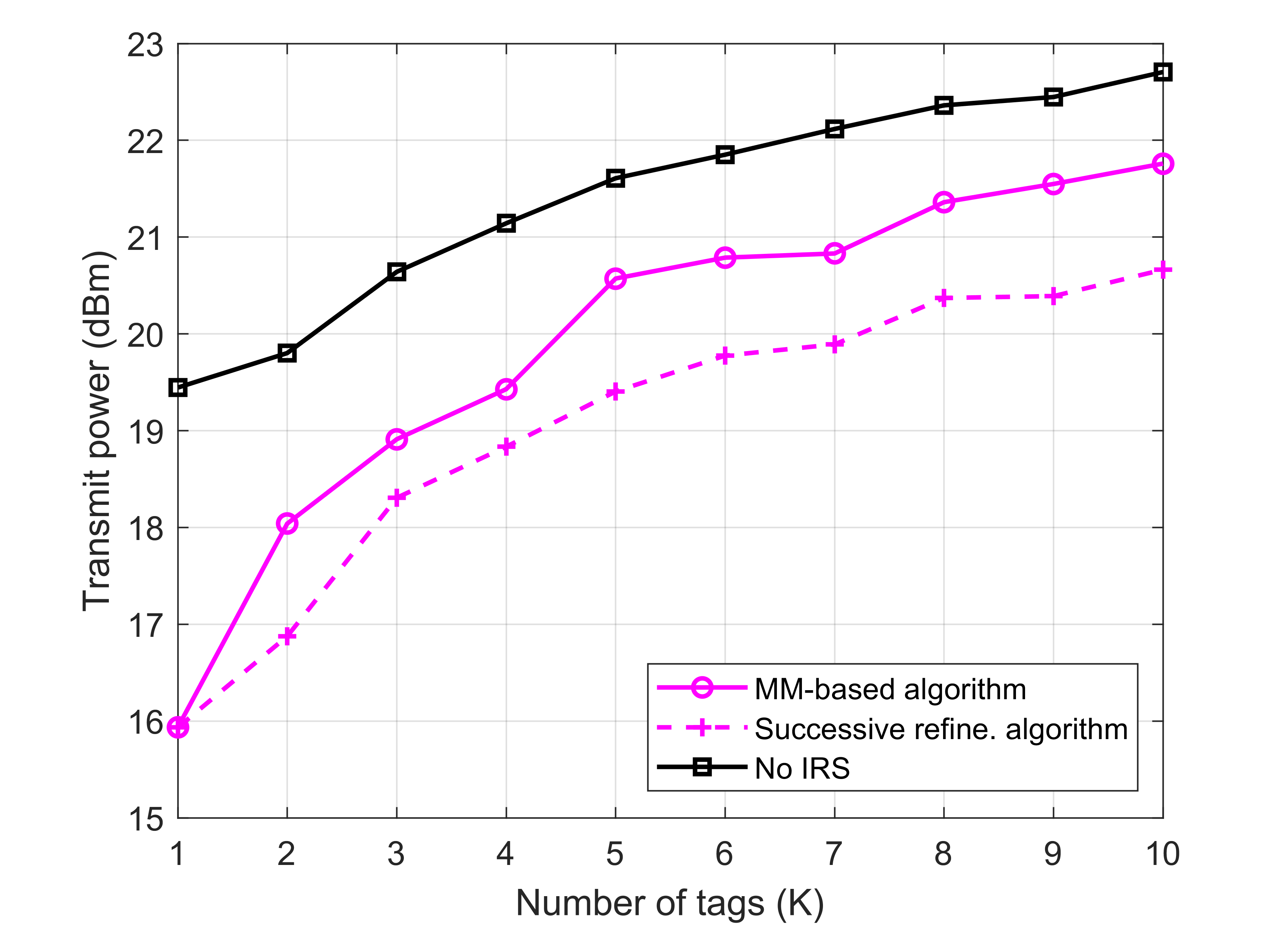}
		\caption{No circuit power constraint}
	\end{subfigure} \hfill
	\begin{subfigure}{0.495\textwidth}
		\centering
		\includegraphics[width=3.5in]{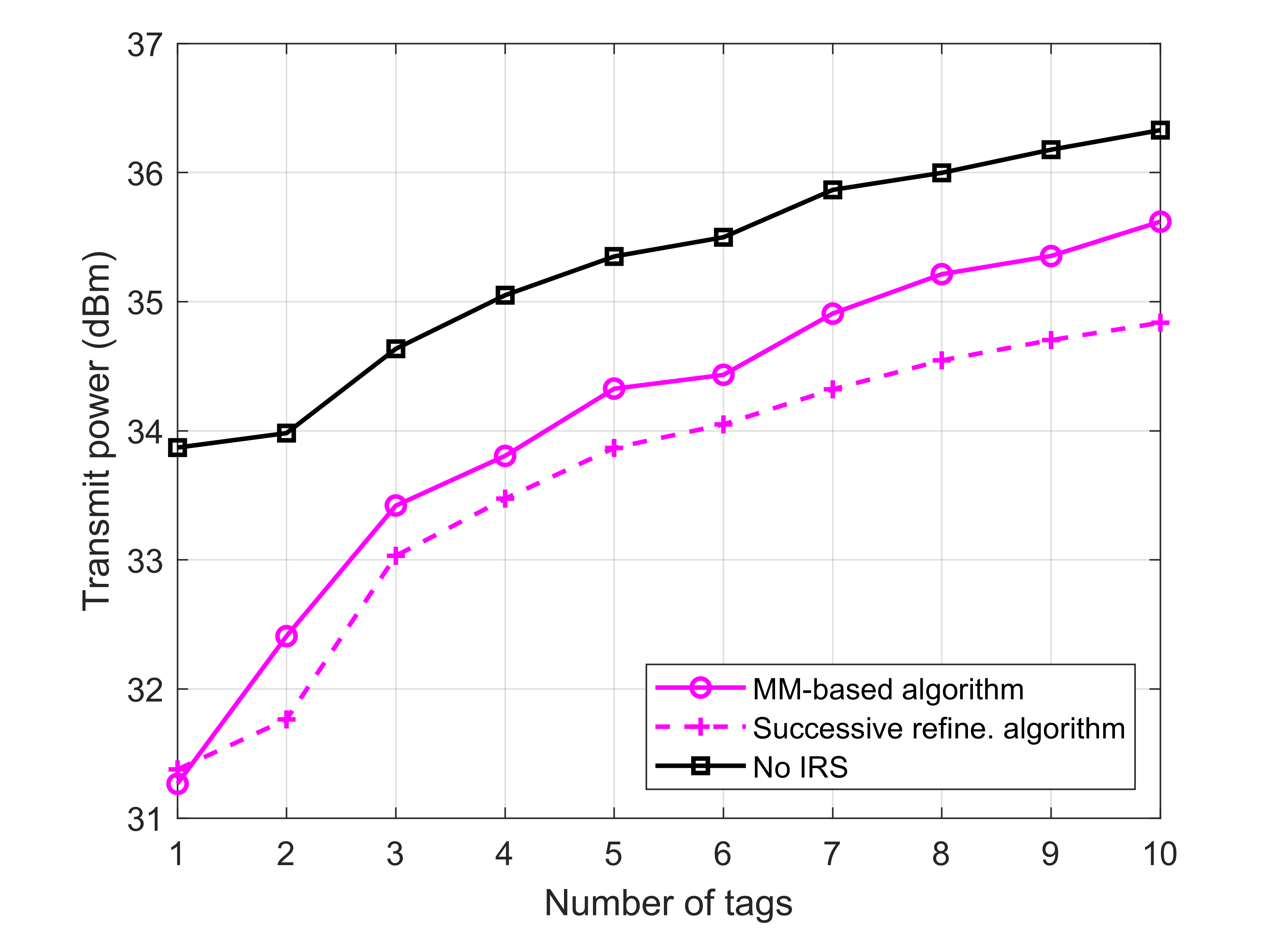}
		\caption{Nonzero circuit power constraint ($\xi = -22$ dBm)}
	\end{subfigure}
	\caption{Effect of multiple tags on the transmit power at the CE.}
	\label{fig:multiUserNumTags}
\end{figure}

\textit{2) Effect of Number of Tags:} Fig. \ref{fig:multiUserNumTags} shows the effects of increasing the number of tags served on the transmit power, where tags are semi-passive and passive. One may observe that the transmit power reduction decreases with more tags, from roughly $3.5$ dB in a single-tag scenario in Fig. \ref{fig:multiUserNumTags}(a) to around $2$ dB for $K = 10$. Nonetheless, this reduction is still significant, given that the algorithms attempt to obtain the most favorable phase shift matrix to balance the performance of all tags, not all of whose channels may be favorably aligned. Moreover, the MM-based approach is outperformed by the SR scheme for $K \geq 2$ (albeit by a smaller margin in the circuit power constrained case), highlighting its declining performance as more constraints (proportional to the number of tags) are imposed and approximated with quadratic forms. The latter algorithm achieves a similar effect compared to randomization in \textbf{Algorithm 2} without performing the randomization or needing to factor in the structure of the SNR and circuit constraints. As a result, the SR algorithm achieves more favorable objective values for large $K$. Overall, the effective transmit power per tag decreases with larger $K$.

Fig. \ref{fig:multiUserNumTags}(b) presents the case where a circuit power consumption of $\xi = -22$ dBm is required. A significant increase in the base transmit power is observed for both algorithms, similar to the single-tag system with circuit constraint. In addition, the relative reduction in transmit power compared to the semi-passive tag case is slightly diminished. This is in part due to the circuit constraint in the optimization problem causing the phase shifts to favor the combined CE-tag link, which may not be favorably aligned with the combined tag-reader link. For both semi-passive and passive tag scenarios, we observe that lower transmit power is incurred on average per tag as the number of tags increases.


\section{Conclusion}
In this paper, we studied the novel integration of IRS into a bistatic BackCom system, and presented its corresponding signal model. A first attempt at obtaining solutions for the transmit power minimization problem at the CE was undertaken using the AO and MM techniques, to address the highly nonconvex nature of the problem brought about by the twice-reflection of the signal traveling from the CE to the reader; in addition, low-complexity successive refinement algorithms were also proposed for the phase shift optimization. Our findings suggest that the introduction of an IRS, even if moderately sized, has a considerable effect in reducing the CE transmit power compared to a non-IRS-aided system. Moreover, for multi-tag systems, the transmit power was found to scale favorably in terms of power consumption per tag, as the number of tags is increased. As an exploratory work into IRS-aided BackCom systems, there are many interesting avenues for future research. The transmit power minimization problem studied in this paper could be reformulated as an energy efficiency maximization problem, to investigate the energy usage-throughput trade-off. Joint optimization could be considered for systems with multiple PBs; and the case with multiantenna tags also warrants further study.




%
%


\ifCLASSOPTIONcaptionsoff
  \newpage
\fi

\bibliographystyle{ieeetran}
\bibliography{IEEEabrv,irs_ref}

\end{document}